\documentclass{aa}  

\usepackage{graphicx}
\usepackage{txfonts}
\usepackage{xcolor}

\begin{document} 

   \title{Stellar populations from H-band VLT spectroscopy in a sample of seven active galaxies
   ~\thanks{Based on observations collected at the European Organisation for Astronomical Research in the Southern Hemisphere, Chile, under programs 63.A-0366 and 71.B-0228A, and HST archive 
images. }}

   \author{
I.~M\'arquez\inst{1}
\and
C.~Boisson\inst{2} 
\and
M.~Joly\inst{2}
\and
D.~Pelat\inst{2}
\and
F.~Durret\inst{3}
}

\institute{
Instituto de Astrof\'isica de Andaluc\'ia, CSIC, Glorieta de la Astronom\'ia s/n, 18008, Granada, Spain
\and
LUX, Observatoire de Paris, Universit\'e PSL, Sorbonne Universit\'e, CNRS, 92190 Meudon, France
\and
Sorbonne Universit\'e, CNRS, UMR 7095, Institut d'Astrophysique de Paris, 98bis Bd Arago, 75014, Paris, France
}

\date{Received ; Draft printed: \today}
 
  \abstract
   { The relationship between an active galactic nucleus (AGN) and its host galaxy is still 
   far from being  understood. Properties of the host galaxies of Seyfert nuclei, such as luminosity concentration, morphological type, metallicity, and age of the stellar population, are expected to be related with nuclear activity -- either at the epoch of galaxy formation or at the present day via feeding of the central black hole.
}
   { In this paper, we investigate whether stellar ages and metallicities are linked to the activity within the nucleus
   in a sample of AGNs of various types. 
   }
   { Our sample includes seven AGNs, from Seyfert~1 to LINERs, observed with VLT/ISAAC and VLT/SINFONI. Based on an inverse method using a stellar library, we analysed H band infrared spectra,  in a wavelength region devoid of emission lines,  at a spectral resolution of $\approx$3000, in the few central 100~pc.  Hubble Space Telescope images were used to visualise the regions defined in each galaxy.  }
   { For each galaxy, we give the results of the spectral synthesis, in particular the percentages of the stellar, power law, and black-body continua, and the percentages  of various stellar types that  account for the stellar lines. }
   { Out of the seven galaxies, three show strong and recent star formation in the inner 100~pc, while no star formation is detected in the three genuine Seyfert~2 galaxies. Beyond a radius of 100~pc, all show more or less recent star formation. 
  Moreover, we can  conclude that the star formation history of the inner nucleus is highly heterogeneous. }

   \keywords{Galaxies: active. }

   \maketitle

\section{Introduction} \label{sec:intro}

An important yet unsolved issue for understanding the Seyfert phenomenon is the relation between the host galaxy and the properties of activity within the nucleus. It is now accepted that nuclear star formation and central engine activity can coexist in the inner regions of the active galactic nuclei (AGNs) host galaxy, as derived from the strong relation linking the mass of super-massive black holes with the velocity dispersion of their host galaxy bulge \citep[e.g. ][]{Gebhardt2000,  Kormendy2013}. The accelerated star formation caused by dynamic instabilities that trigger and/or fuel nuclear activity could result in an overabundance of giant, supergiant, and supermetal-rich (SMR) stars \citep[e.g. ][and references therein]{Scoville2023}.

Stellar population syntheses of the cores of AGNs provide information on stellar masses,  mean stellar ages, abundances, and dust content, as well as insight on their star-formation history. Such information is derived from spatially unresolved galaxy spectra and thus reflect global averages over properties that are known to vary across 
the galaxies, but still provide valuable insights.  Observations of the cores of nearby active galaxies allow the complex star formation history in active galaxies to be exhibited \citep[e.g. ][]{Terlevich1990, StorchiB2001, Diniz+17, Mallmann+18, Dametto+19}. Studies in the optical and infrared (IR) domains have indicated that young or intermediate age populations can contribute significantly to AGN central spectral emission \citep[e.g. ][and references therein]{Oliva+99,Boisson+00,Chapman2000,Boisson+04,GarciaBernete+15, Thomas2018, Burtscher+21, Dahmer+22, Riffel+24}.

To study the properties of the stellar population in the inner few tens of parsecs, all studies face the problem of disentangling the stellar population from the other components: the continuum emission by the AGN, the emission from heated dust, and galactic extinction.  Contrary to the optical spectral range, near-IR allows one to access to regions highly obscured by dust, and exhibits many stellar features that permit one to constrain the average spectral type, luminosity class, and metallicity markers characteristic of a large variety of stellar populations \citep[e.g. ][]{Origlia1993,Dallier1996,Ivanov2004,Stephens2004}.
Since part of the near-IR H-band is free of the strongest emission lines due to the AGN, it is a well-suited domain for the determination of the stellar content of central regions of Seyfert galaxies, including Seyfert~1s, as long as the spectral resolution is high enough  \citep[e.g.][]{Dallier1996,Fremaux+06,Cesetti2009}. 

In this context, we quantify preliminary results of \cite{Boisson+02} with  R$\approx$ 3000 spectral resolution in the H-band range ($\sim$ 1.57 to 1.64 $\mu$m) for a sample of Seyfert galaxies. Our data encompass the core to a few hundreds of parsecs, in order to highlight possible population gradients and locations of more or less recent starburst periods. 

In Section \ref{sec:data} of the paper, we briefly present the sample of seven nearby AGNs observed with the VLT and summarise the data reduction. The spectral population synthesis code and its implementation are presented in Section \ref{sec:synthesis}. Results for individual AGN  are presented in Section \ref{sec:S_fit} and discussed in a global context in Section \ref{sec:discussion}.

\section{The data}
\label{sec:data}

To study the differential stellar population in the cores of AGNs, we assembled a sample of seven sources typical of their class of activity, three Seyfert 1 galaxies (MCG-6-30-15, NGC~2992 and  NGC~3783) and four Seyfert 2s (NGC~2110, NGC~3185, NGC~6221 and NGC~7582). Five galaxies observed with the VLT/ISAAC spectrograph, were previously presented in  \cite{Boisson+02} and are complemented by two more galaxies (this paper) observed with the same equipment. We combined these data with VLT/SINFONI  spectra kindly provided by Sandra Raimundo for MCG-6-30-15 \citep{Raimundo+13}.  

Hubble Space Telescope (HST) archival images were used to visualise the different regions as selected for each  galaxy. Informations on the seven galaxies of this study are given in Table~\ref{tab:listgal}.
In this paper we assume H$_0$ = 70 km s$^{-1}$ Mpc$^{-1}$, $\Omega_m$=0.3, $\Omega_{\Lambda}$=0.7.

\begin{table*}
  \caption{Galaxies observed with Isaac. } 
\begin{tabular}{lccllccrlcc}
\hline
\hline
Name & Sy  &Morphology& RA        & DEC     & z   & E(B-V) & PA    & HST   
& D  &
scale \\
     & type     &&(J2000.0)  &(J2000.0)&     &        & ($\deg$)& filter& (Mpc) 
& (pc/")\\ 
\hline
MCG-6-30-15 & 1& E/S0&13h35m53.7s &-34d17m44s & 0.00775 & 0.06  & 86 &F547M & 31 & 150\\
NGC~2110 &2&E3/S0& 05h52m11.4s & -07d27m22s & 0.00779&0.37 &$-25$  &fr680p15& 32 & 157 \\
NGC~2992 &1/2&Sa& 09h45m42.0s & -14d19m35s & 0.00771&0.06 & 55    &F196N &  31 & 150 \\
NGC~3185 &2&SBa& 10h17m38.5s & +21d41m18s & 0.00406&0.03 & 0     &F814W & 16 & 78 \\
NGC~3783 &1&SBa& 11h39m01.7s & -37d44m19s & 0.00973&0.12 &$-77$  &F160W &  40 & 189 \\
NGC~6221 &2&SBbc& 16h52m46.1s & -59d13m07s & 0.00500&0.17 &30     &F160W &  19 & 92\\
NGC~7582 &2&SBab& 23h18m23.5s & -42d22m14s & 0.00525&0.01  & 0     &F606W &  22 & 107\\
\hline
\end{tabular} 
The columns are: 1)~name,
    2)~Seyfert type, 3)~morphological type, 4) and 5)~coordinates, 6)~redshift, 
7)~Galactic
    E(B-V), 8)~slit position angle, 9)~HST filter, 10)~distance, 11)~scale in parsecs 
per arcsecond. Galactic E(B-V) values are from \cite{SchlaflyFinkbeiner11} as given by the Dust Extinction tool provided by IRSA (https://irsa.ipac.caltech.edu/applications/DUST/).
\label{tab:listgal}
\end{table*}

\subsection{VLT/ISAAC data}
\label{sec:isaac}

Using the ISAAC spectrograph mounted on the VLT/ANTU telescope, we obtained, in service mode, long-slit medium resolution (R$\approx$3300) H-band spectroscopy for five galaxies  as published in  \cite{Boisson+02} and for two more observed in 2003, reported in this paper\footnote{Programs ESO 63.A-0366 and 71.B-0228A}.
With a spectral sampling of 0.79 \AA, the 1~arcsec width slit gives a spectral resolution of 4.5 \AA\ full width at half maximum (FWHM). The spatial sampling is 0.147"/px, and the average seeing was between 0.6 and 1 arcsec. 
 
The slits were always centred on the IR galaxy nucleus. 
Their position angles are provided in Table \ref{tab:listgal}.

The data reduction is fully described in \cite{Boisson+02},
including correction for electrical ghosts, flat-field,
distortions along and perpendicular to the slit, and
wavelength calibration. Early type stellar templates were used for the 
correction of telluric absorption. The spectra were extracted by using
logarithmic flux profiles to define a nucleus and regions around
it. All of the spectra displayed are shifted to the rest wavelength.
Whenever spectra on each side of the nucleus were very similar, a pseudo  
ring spectrum was created by summing both individual spectra, allowing for a better signal to noise ratio.

\subsection{VLT/SINFONI data}
\label{sec:sinfoni}

Data from the near-IR integral field  SINFONI spectrograph on the VLT have a  
spectral resolution of R$\approx$3000, the mean value for the instrumental 
broadening being 3.2 \AA ~at 1.6187 $\mu$m. The pixel scale of the final 
reduced cube is 0.05 arcsec, and the point spread function is well modelled by a 
double Gaussian with FHWM of 0.07 and 0.2 arcsec for the peak and the wing 
components, respectively. The data reduction process and the spectral 
extraction procedures are provided in \cite{Raimundo+13}. In the following, 
we study the four regions as marked in their Fig.7 and an integrated spectrum of 
a ring over a radius of $\approx$ 1.1 arcsec, excluding the central 0.3 x 0.4 
arcsec$^2$ (see Fig. 6 in \citep{Raimundo+13}).

\subsection{HST images}
\label{sec:hst}

Archival WFPC2 HST images are available for all the galaxies (see the
HST filters in Table \ref{tab:listgal}). 
The reddest high quality images were  used. In order to enhance morphological features in the innermost few arcseconds of the galaxies, we created sharp-divided images by dividing the original images by 
their smoothed counterparts \citep{Marquez03}. This enhanced any 
structural inhomogeneities that could affect the stellar populations, such as a dust 
component. 

The spectral regions extracted from ISAAC and SINFONI observations were drawn on 
the images, and the slit width is 1\arcsec. North is up and east to the left.

\section{Stellar population synthesis}
\label{sec:synthesis}
The analysis of the stellar populations inside unresolved cores of active galaxies has been the aim of several studies for some time now. All studies face the problem of disentangling the stellar population from all the other components: the continuum emission by the AGN, the emission from the heated dust, and galactic extinction. 

The near-IR range is a very useful domain to study
the stellar content of active galaxies. In particular, the H-band is
well suited, as the non-stellar contribution (mainly reprocessed
nuclear emission by dust) is much less than in the K-band window.
The H-band is also the region in which emission from 
cool stars peaks, and the spectrum is rich in luminosity sensitive stellar absorption lines,
leading to the age of the stellar population 
\citep[e.g. ][]{Dallier1996, Fremaux+06}.  
Extinction by dust is lower there than for the shorter wavelength J-band. 

Since part of the IR H-band is free of the strongest AGN emission lines, it may allow for the determination of stellar populations in the very internal regions of Seyfert galaxies in the presence of a diluting continuum, as is the case for Seyfert 1s. In particular the spectral range considered in this paper, covering  1.57 to 1.64 $\mu$m,  includes many metallic stellar features (FeI, MgI, SiI, CaI, NiI, CO, and OH). The spectral variations of the stellar component are generally complex, and the age and the metallicity of the stars may have similar influences on the integrated spectrum. Therefore, it is crucial to use as many observables as possible to decrease the degeneracy. 
 Indeed the near-IR CaII triplet and the IR CO and Si indices can be strong not only in supergiant stars but also in metal-rich giants; colour gradients can be caused either
by changing stellar populations or by dust effects. 

There are essentially two different approaches to differential population synthesis. The 'direct' approach, also called the 'evolutive synthesis method' \citep[e.g.][]{Tinsley1972, Bruzual03, leborgne04,Vazquez05}, compares the spectrum at each time step of an evolving stellar population with the spectrum of the studied galaxy. The other method aims to extract the stellar components from the spectrum of the astrophysical object through a mathematical inversion of the information content of the observed spectrum. This so-called inverse method uses building blocks of either stellar spectra \citep[e.g. ][]{Faber_1972, Pickles_1985, Pelat97, Pelat98, Moultaka_Pelat2000, Fremaux+07} or simple stellar populations \citep[SSPs; e.g.][]{Bica1988, Vergely2002, Cid+05, Gomes2017, Wilkinson+17,Cid2018}. 
In this paper, we use an inverse method, based on the minimisation of a quadratic form subject to constraints in the form of inequalities and equalities, developed and thoroughly tested by \cite{Fremaux+07}.

Taking the flux received in each spectral pixel as the observable quantity to fit increases the number of observables, allowing for a greater sampling for the stellar composition of the studied galaxies, and alleviating the difficulties to determine the level and shape of the continuum. The specificity of the method is to take the non stellar parameters (reddening, dust emission, and AGN non-stellar continuum) directly into account in the synthetic distance to be minimised. These non-stellar parameters introduce non-linearity into the objective function, expressed as a quadratic distance to minimise. The lower this value, the closer from the observation the synthetic spectrum. 

Due to degeneracy of these components, mainly in low S/N ($<$ 40), the non-stellar parameters are bound to values commonly deduced from astrophysical considerations. Dust grains with a temperature below 200 K mainly contribute to the flux spectrum farther in the IR, so their contribution in our wavelength range is too small to be detected.  A temperature of 1500~K is the maximum possible for dust grains as it corresponds to the sublimation temperature of the most refractory dust material. Most of the active galaxies have an intrinsic reddening E(B-V) smaller than 3. For each galaxy the minimum allowed galactic extinction was fixed to the Galactic value on the line of sight \citep{SchlaflyFinkbeiner11}. The AGN continuum power law index $\alpha$ is the most difficult non stellar parameter to estimate. Indeed, in our very small wavelength domain, variations of $\alpha$ only induce a very weak variation in the slope of the fitted continuum, as explained in \cite{Fremaux+07}.  We therefore fixed $\alpha$ to the 
typical value of 1.5 for AGNs\footnote{the power law is defined by $\nu ^{-\alpha}$ } \citep{VandenBerk+01}. However, these bounds can easily be modified for the synthesis of individual objects.

Such a method is inherently prone to errors due to  the presence  of noise in the data and potential degeneracies inherent to the studied problem. To assess the impact of these errors on the estimated parameters, we performed a series of Monte Carlo simulations. Gaussian random noise with zero mean and a standard deviation of 10\% of the observed galactic spectrum value was added to each pixel of the spectrum. This corresponds approximately to the signal-to-noise ratio of the data. For each object, 300 independent Monte Carlo trials were conducted.
To mitigate the influence of outliers, 38\% of the highest and 38\% of the lowest values for each parameter were excluded. This percentage corresponds to the optimal truncated estimator for the worst-case scenario, that is noise with an undefined mean (Cauchy noise). The standard deviation of the remaining values provides the error bars for the parameters obtained via the minimisation method.

Whatever the approach in stellar syntheses, a spectral database is needed to build a synthetic galactic spectrum. The stellar libraries come in two flavours, either built with theoretical or empirical spectra. Though theoretical libraries could allow for a wide range and combination of stellar parameters, it requires exhaustive model atmospheres that depend on atomic and molecular line opacity data bases not always available.

As demonstrated by \cite{Fremaux+06}, the comparison between observed and computed spectra deteriorates for decreasing temperature, where several molecular as well as metallic absorption lines are still missing in atmospheric models. Working in the IR H-band, we thus use an empirical stellar library. It should be noted that each individual spectral type in a library encompasses a significantly large domain of temperature, gravity, and metallicity, because one cannot have as many types of stars in the database as those forming a galaxy. 

To match the spectral resolution of the galaxy observations, R$\approx$3000, the stellar library used in the present study is a sub-sample from H-band observations by \cite{Meyer+98} complemented by ISAAC stellar spectra, most of them from supermetallic stars \cite[see][]{Boisson+02}. 
Unfortunately there are not that many data bases in the IR; they are usually of much lower spectral resolution \citep[e.g.][]{Rayner_2009, Villaume_2017}, or do not have any supermetallic stars \cite[e.g. ][]{Verro_2022}.

Also, the finite  spectral resolution of observations imposes a limitation in the extent of the library due to possible degeneracy among stellar spectra. 
To avoid multiple degenerate solutions for the minimisation problem, the library is composed of 30 stars of all luminosity classes and of spectral types from A0 to M4, chosen to describe at best the HR diagram \cite[see discussion in][]{Fremaux+07}.  
The code includes physical constraints  on the relative number of born stars of the different mass groups, here in the case of a standard Initial Mass Function \cite[see a discussion in][]{Moultaka2004}. Table~\ref{tab:masses} gives the spectral type, absolute visible magnitude, temperature, luminosity, mass, and age of each of the stellar types 
considered together with the name of a typical star. When known, the metallicity ([Fe/H]) of this star is provided with the associated 
reference. Although we do not intend to give the exact metallicity of each stellar component of a galaxy, we should be able to distinguish qualitatively a metal-rich  
from a solar metallicity population. In the following we refer to super metal-rich stars as SMR.

In the text, we define as young, intermediate and old stellar populations those 
with stars with ages $\leq$ 10$^{7}$~yrs (supergiant stars), around 10$^{8}$~yrs (A0-F2V), and older than 10$^{9}$~yrs (dwarfs and giants) respectively. The supergiant stars are indeed used as a proxy for a young population since no stars hotter than A type can be considered; the contribution of O and B type stars is negligible in the IR, compared to cold or intermediate 
stars (A, F spectral types).
For more details on the spectral fitting code and database see \cite{Fremaux+07}.

\begin{table*}
\caption{Stellar parameters. }
\begin{tabular}{lllrrrlrl}
\hline
\hline
Name & Spectral  & M$_V$ & T(K)  & L/L\sun  & M(M\sun)& Lifetime & [Fe/H] & Ref.\\
 &Type & & & & & (yrs) & & \\
\hline    
HD172167  &A0V            & 0.5   & 9500  & 54       & 3.0     &  2.7$\times$10$^{8}$   & & \\        
HD48501   &F2V            & 3.6   & 7100  & 2.9      & 1.5     & 3.9$\times$10$^8$     &  0.01& BT\\     
HD112164  &rG1V           & 4.5   & 5900  &  1.0     & 1.0     &6.0$\times$10$^9$     &  0.24& E\\       
HD10307   &G1.5V          & $>$4  & 5880  & 1.0      & 1.0     & 6.2$\times$10$^9$     &  0.0 & L\\      
HD106116  &rG4V           & 4.9   & 5700  & 1.0      & 1.0     & 6.9$\times$10$^9$    &  0.15& An\\ 
HD185144  &wK0V           & 6.0   & 5300  & 0.42     & 0.8     & 6.4$\times$10$^{9}$   & -0.18& R\\      
HD39715   &K3V            & 6.65  & 4740  & 0.32     & 0.7     & 1.1$\times$10$^{10}$  & -0.04& VF\\     
HD131977  &K4V            & 7.2   & 4600  & 0.2      & 0.66    & 3.5$\times$10$^{10}$  &  0.05& L\\      
HD201091  &wK5V           & 6.0   & 4400  &0.32      & 0.7     & 3.2$\times$10$^{10}$  & -0.35& KV\\     
HD79210   &M0V            & 9.0   & 4000  & 0.09     & 0.5     & $>$3$\times$10$^{10}$  & 0.0 & M\\     
HD36395   &rM1V           & 9.3   &3700   & 0.07     & 0.45    & $>$3$\times$10$^{10}$  & 0.49& M\\     
HD119850  &wM1.5V         & 9.3   &3700   & 0.07     & 0.45    & $>$3$\times$10$^{10}$ & -0.3& M\\     
HD173739  &wM3V           & 11.5  &3700   &0.02      &0.25     & $>$3$\times$10$^{10}$  & -0.29& M\\     
HD27397   &F0IV           & 2.7   & 7000  & 6.5      & 1.3     & 2.1$\times$10$^{9}$   &  &\\       
HD115604  &rF3III         & 1.6   & 6800  & 17       & 1.5-1.6 & 3.0$\times$10$^{9}$   &  0.18& H\\      
HD220657  &F8III          & 1.2   & 5800  & 30       & 1.2     & 6.0$\times$10$^{9}$   &  &\\       
HD107950  &G6III          & 0.85  & 5000  & 40       & 1.0     & 9.3$\times$10$^{9}$   & 0.08& ASF \\    
HD206952  &rK1III         & 0.6   & 4700  & 70       & 1.1     & 6.0$\times$10$^{9}$   &  0.2& P\\       
HD3627    &K3III          & 0.5   & 4300  & 110      & 1.10     & 6.0$\times$10$^{9}$   &  0.04& MLSF\\  
HD3346    &K5-M0III       & -0.2  &3590   & 330      & 1.1     & 6.0$\times$10$^{9}$   &  &\\
HD102212  &wM1IIIab       & -0.5  & 3800  & 430      & 1.25    & 6.0$\times$10$^{9}$   & -0.3& S\\       
HD151732  &M4III          & -0.5  & 3535  & 880      & 1.3     & 4.0$\times$10$^{9}$   &  &\\       
HD19058   &wM4II          & -1.2  & 3500  & 450      & 5       & 1.1$\times$10$^8$     & -0.15& SL\\     
HD36673   &F0Ib           & -6.6  & 7400  & 30000    & 12      & 1.8$\times$10$^{7}$   &  0.04& V\\      
HD20902   &F5Ib           & -6.6  & 7000  & 30000    & 12      & 1.8$\times$10$^{7}$   & -0.05& P\\      
HD139717  &rF8Ib      &  -6.5   & 6500  & 30000    & 9       & 2.5$\times$10$^{7}$   &  0.15& Ad\\
HD204867  &rG0Ib          & -6.4  &5500   &30000     & 10      & 1.4$\times$10$^{7}$   &  0.10& KV\\     
HD210745  &rK2I           & -5.95 & 4300  &30000     &12       & 1.7$\times$10$^{7}$   &  0.17& KV\\     
HD201251  &rK4I            &  -5.8 &  4000 & 33000    & 12      & 1.8$\times$10$^{7}$  & 0.12 & Lu\\  
HD39801   &M2I           &  -6.0 &  3500 & 63000    & 18
& 8.5$\times$10$^{6}$ &0.01 & P\\ 
\hline     
\end{tabular}  
\\
r subscript for SMR; w  subscript for below solar metallicity\\
References: An: \cite{Andrievsky+13}; Ad: \cite{Adibekyan+12}; ASF: \cite{Afsar+12}; BT: \cite{Boesgaard+87}; E: \cite{Edvardsson+93}; H: \cite{Hauck+85}; 
KV: \cite{Koleva+12}; L: \cite{Lee+11}; Lu: \cite{Luck14}; M: \cite{Mann+15}; MLSF: \cite{Massarotti+08}; P: \cite{Prugniel+11}; R: \cite{Ramirez+13}; S: \cite{Sheffield+12}; SL: \cite{Smith+86}; V: \cite{Venn95}; VF: \cite{Valenti+05}. 
\label{tab:masses}
\end{table*}

\section{Spectral fits} 
\label{sec:S_fit}

Synthetic stellar populations are computed for all the individual spectra 
extracted from the galaxy frames, using the stellar database previously 
described.  

The velocity dispersions, $\sigma$, of the galaxies observed with 
ISAAC,  were measured by cross-correlation of the full observed H-band 
spectrum of the galaxy with IR template stars  \cite[see][]{Boisson+00,Boisson+02}.  The resulting $\sigma$ values are given in Tables 3 to 9. The differential velocity dispersion between galaxies and template 
stellar spectra is handled in the synthesis. 

The synthesis results are given in Tables~\ref{tab:mcgres} to \ref{tab:n7582res}, and are displayed in accompanying figures showing the observed
and computed spectra together with the fractional
difference between the model and data. The observed spectra are shown in black, 
normalised by the flux around 1.586 $\mu$m, with the synthesis shown in red. The radius of 
the nuclear region sampled, and the width of the off-nuclear zones are given 
in parsecs. The various parameters listed are: first,  the percentage of stellar (stell), power law (PL) and black-body (BB) emissions, together with the corresponding black-body temperature T(BB), and the intrinsic reddening E(B-V); second,  the stellar composition of the galaxy, together with the percentage  of contribution of each stellar type to the galactic spectrum at the reference wavelength. To ensure readability only stars having a contribution larger than
1\% are listed.

\subsection{MCG-6-30-15}

\begin{table*}
  \caption{Spectral synthesis of MCG-6-30-15. }
\begin{tabular}{rrrrrr|rrr}
\hline
\hline
         & SINFONI  & SINFONI  & SINFONI & SINFONI & SINFONI    & ISAAC & ISAAC    & ISAAC     \\
         & E(1)   & N(2)   & W(3)  & S(4)  &ring & C(nuc)   & E        & W         \\
         & 60-100 pc& 20-60 pc &60-100 pc& 20-60 pc&20-170 pc   & 110 pc&110-330 pc&110-330 pc \\
\hline                     
$\sigma$ km/s& 80          & 80       &80       &80       & 80        & 70    & 80  & 
80  \\
\% stell     & 41$\pm$1   & 23$\pm$1       &  47$\pm$1     & 19$\pm$1      &  39$\pm$1        & 17$\pm$1    & 100$\pm$0 & 
100$\pm$0 \\
\% PL        & 57$\pm$2   & 74$\pm$1       & 52$\pm$1      & 76$\pm$1      &  57$\pm$1        & 83$\pm$1    &   0 & 
  0 \\
\% BB        &  2$\pm1 $   &  3$\pm$0       &  1$\pm$1      &  5$\pm$1     &  4$\pm$0         &  -  &  -  &  
-  \\
T (BB)       & 450$\pm$190  & 250$\pm$10      & 210$\pm$13       & 330$\pm$20   & 290$\pm$6        &    -  &  -   & -
    \\
E(B-V)       & 1.08$\pm$0.2 & -     & 1.27$\pm$0.01     & -    & -       & 0.47$\pm$0.04  & -& 
1.6$\pm$0.1\\
\hline
F2V      &      &      &  16$\pm$3 &      &     8$\pm$3 &       & 49$\pm$1 & 23$\pm$1  \\
rG1V     &     &  14$\pm$10   &      & 13$\pm$8   &  10$\pm$3  &      &    &     \\
rG4V     &      & 22$\pm$4   &      & 9$\pm$10    &      &      &    &     \\
K4V      & 6$\pm$2&  2$\pm$7   &    11$\pm$1  &      & 11$\pm$3   &  14$\pm$3 &    &     \\
M3V      &      &      &    &      &      &        &  17$\pm$1&  28$\pm$1  \\
F8III    & 5$\pm$1   &      &      &      &      &       &    &      \\
K5-M0III &      &      &      &      &      &      &  29$\pm$1&  49$\pm$1   \\
M1IIIab  &      & 4$\pm$13   &      &      &    4$\pm$3 &       &    &      \\
M4III    & 75$\pm$1   & 40$\pm$2   & 37$\pm$1   & 44$\pm$4   &  36$\pm$17  & 68$\pm$3   &    &     \\ 
wM4II    &      &      &      &      &      &      &   2$\pm$1&     \\
rF8Ib    &  8$\pm$1   &      &  9$\pm$2   &      &  7$\pm$1   &      &    &     \\
rK4I     &      &      &      &  2$\pm$1   &     &      &    &     \\
M2I      &   6$\pm$1  & 18$\pm$2   & 27$\pm$1   & 32$\pm$4   & 24$\pm$2   &  18$\pm$3  &  3$\pm$1 &   \\
\hline
\end{tabular}
\\
The numbers given for the main contributing stars are the percentages of the various contributions to the galaxy spectrum at the reference wavelength 1.586~$\mu$m, as in all following tables. 
\label{tab:mcgres}
\end{table*}

\begin{figure}
\centering
\includegraphics[width=7cm]{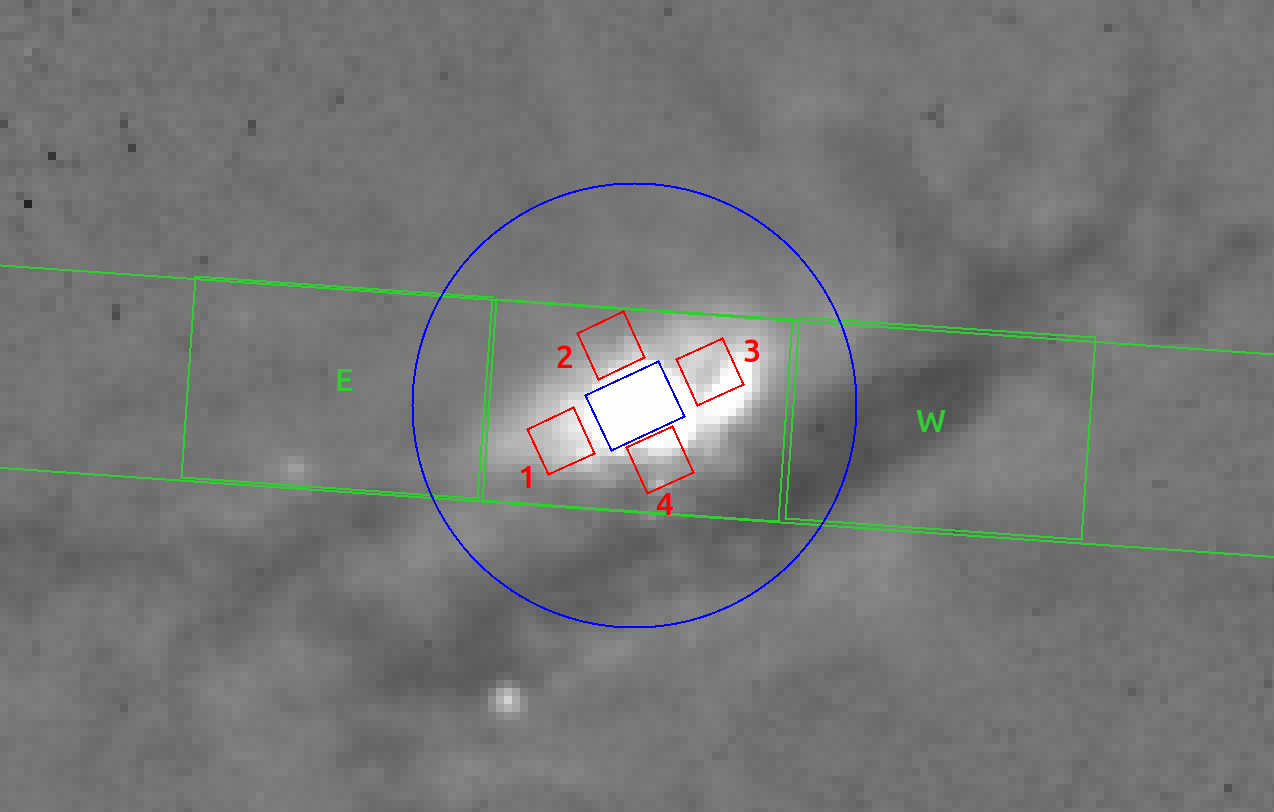}
\caption[]{Sharp-divided optical HST image in the F547M filter of MCG-6-30-15. The region studied by \cite{Raimundo+13} that we call 'pseudo-ring' corresponds to the large blue circle (1.1 arcsec radius) minus the
small blue rectangle, and their regions labelled 1 to 4 are shown in red. The ISAAC slit is superimposed in green, with the centre
and the two regions shown as the central and external rectangles respectively.}
\label{fig:imMCG}
\end{figure}

\begin{figure}
\centering
\includegraphics[width=8.4cm]{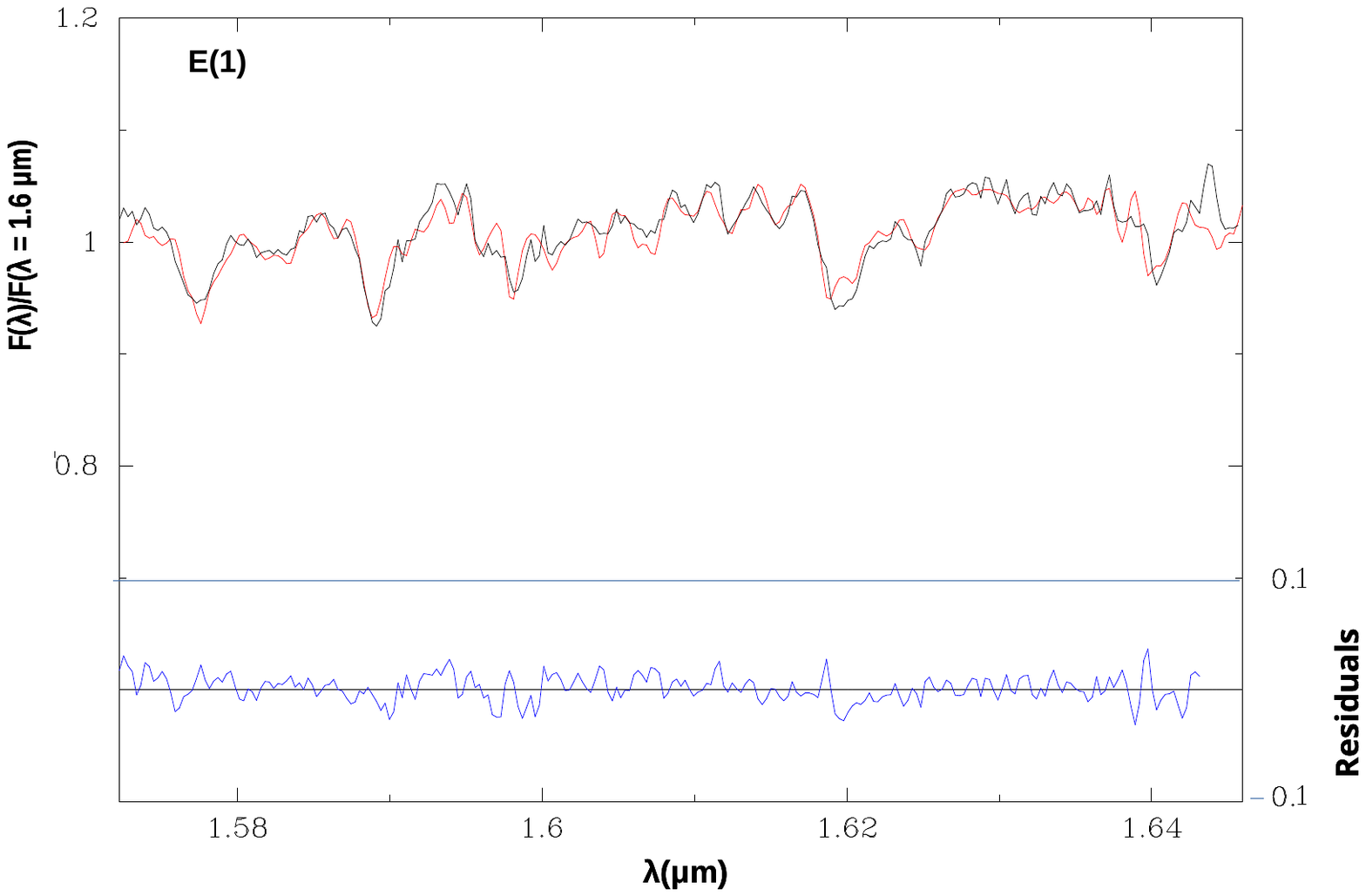}
\includegraphics[width=8.4cm]{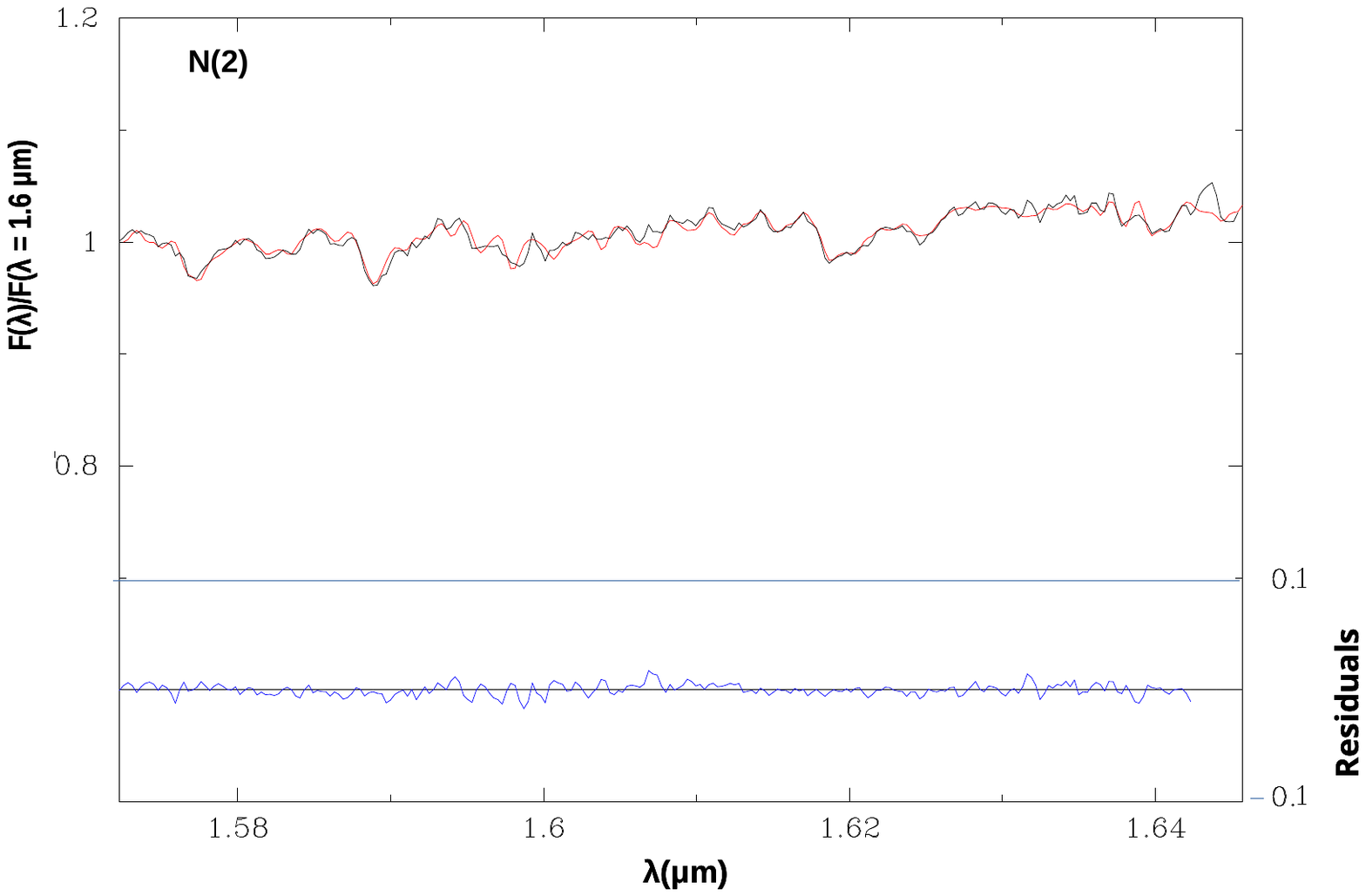}
\includegraphics[width=8.4cm]{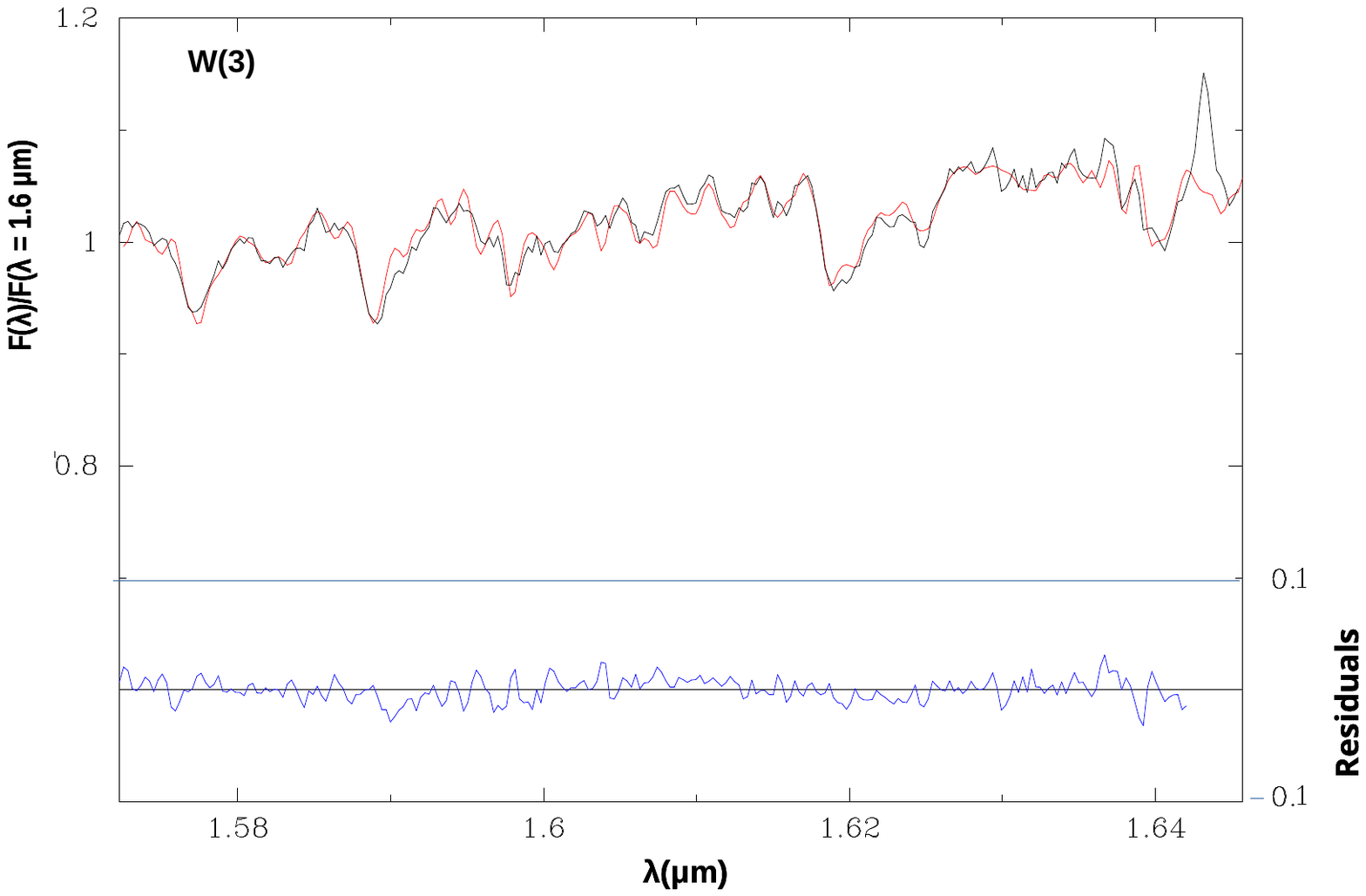}
\includegraphics[width=8.4cm]{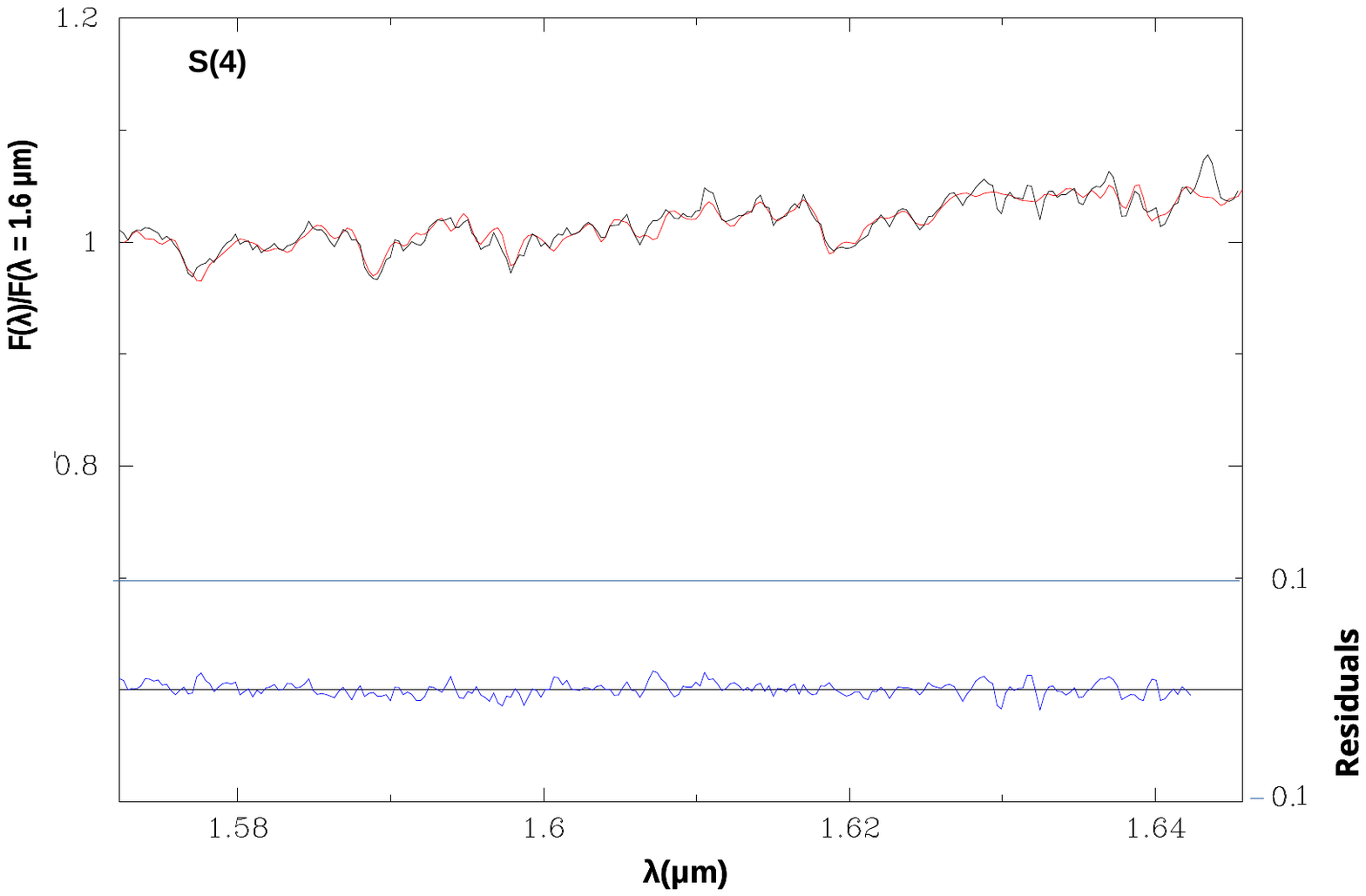}
\caption[]{Stellar population fits to regions 1-4 of  
MCG-6-30-15 (see Fig.~\ref{fig:imMCG} top).}
\label{fig:fitMCG1}
\end{figure}

\begin{figure}
\centering
\includegraphics[width=8.4cm]{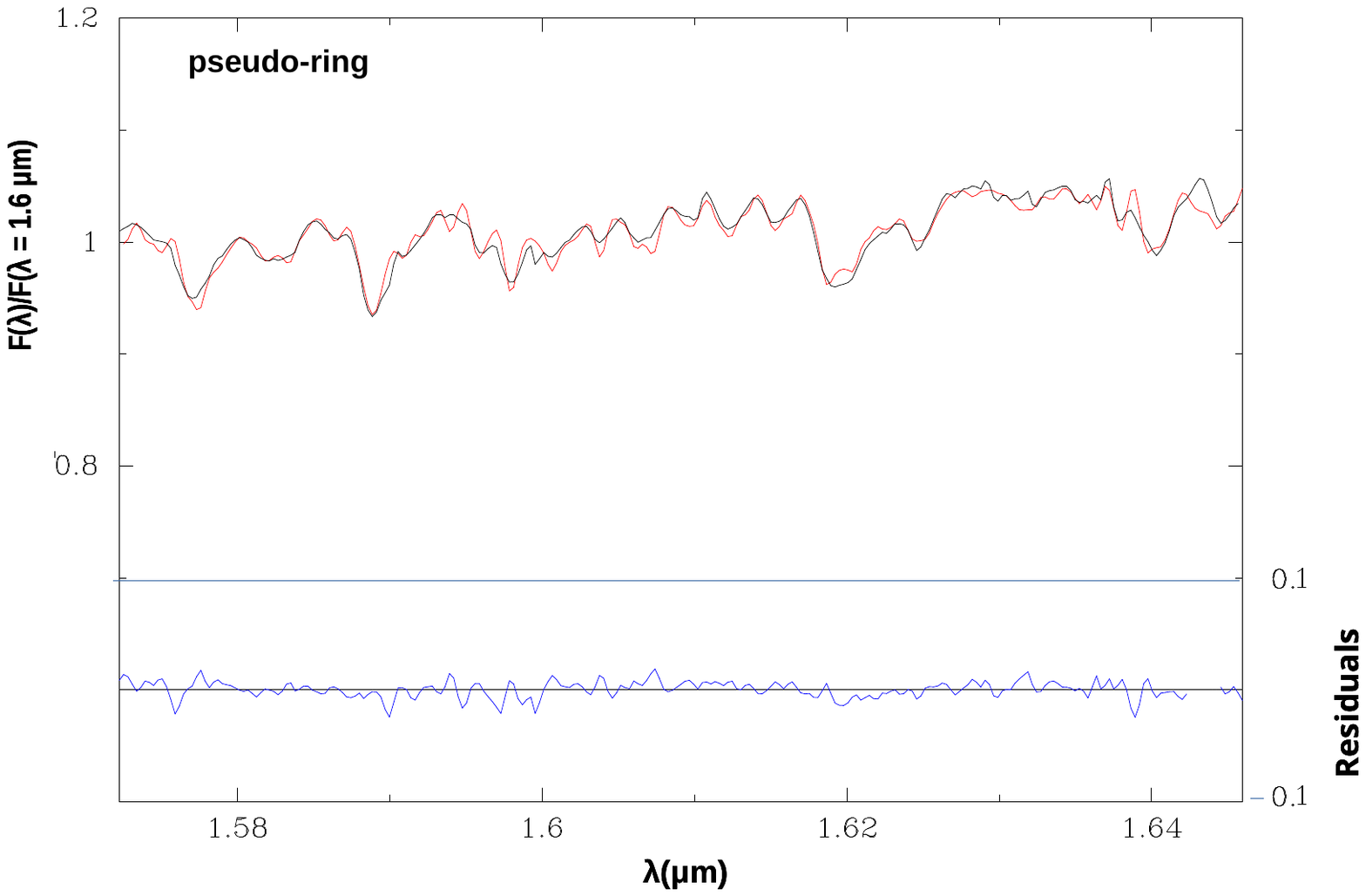}
\caption[]{Stellar population fit to the 'pseudo-ring' of  
MCG-6-30-15 (see Fig.~\ref{fig:imMCG} top).}
\label{fig:fitMCG2}
\end{figure}

\begin{figure}
\centering
\includegraphics[width=8.4cm]{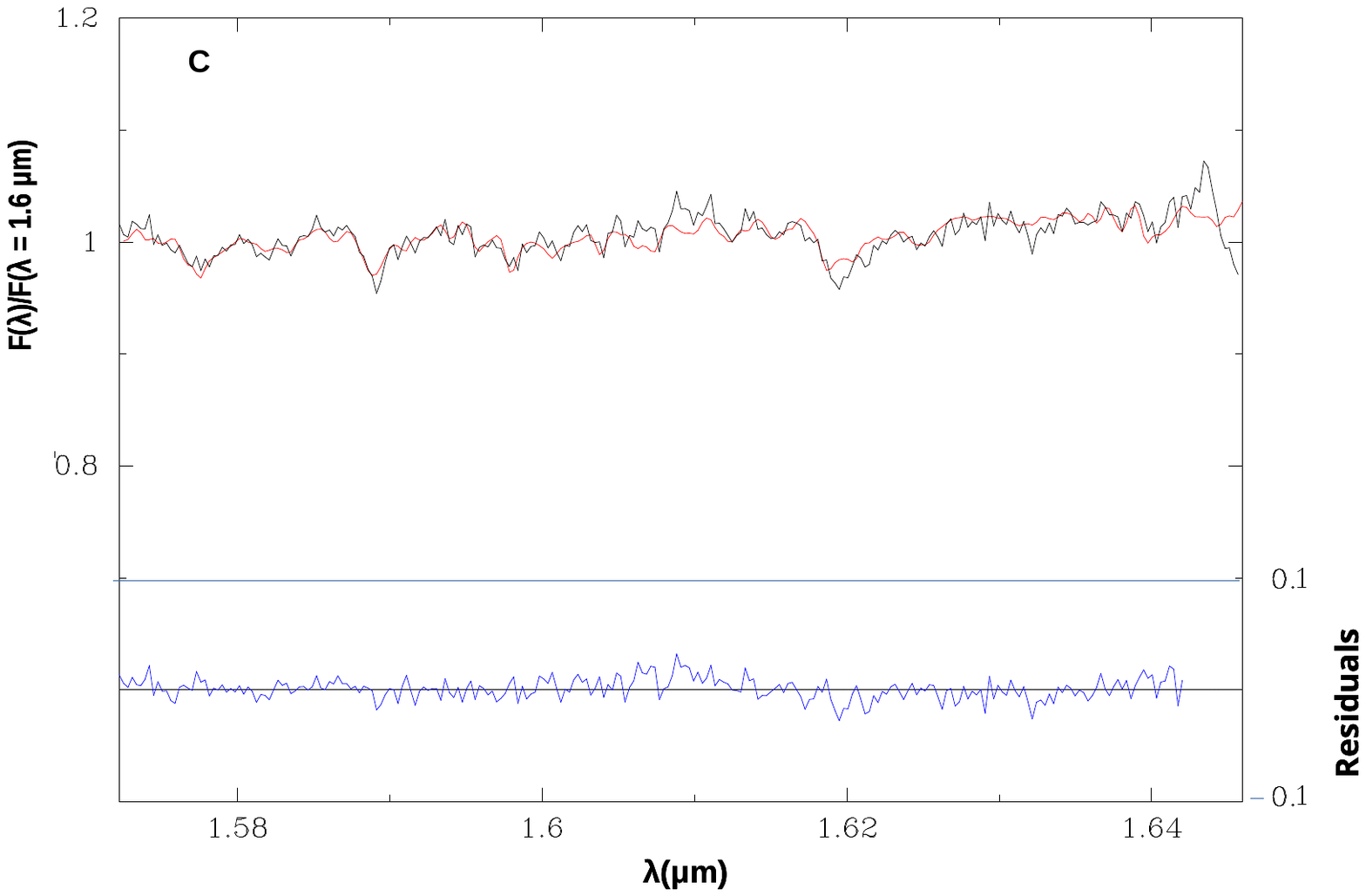}
\includegraphics[width=8.4cm]{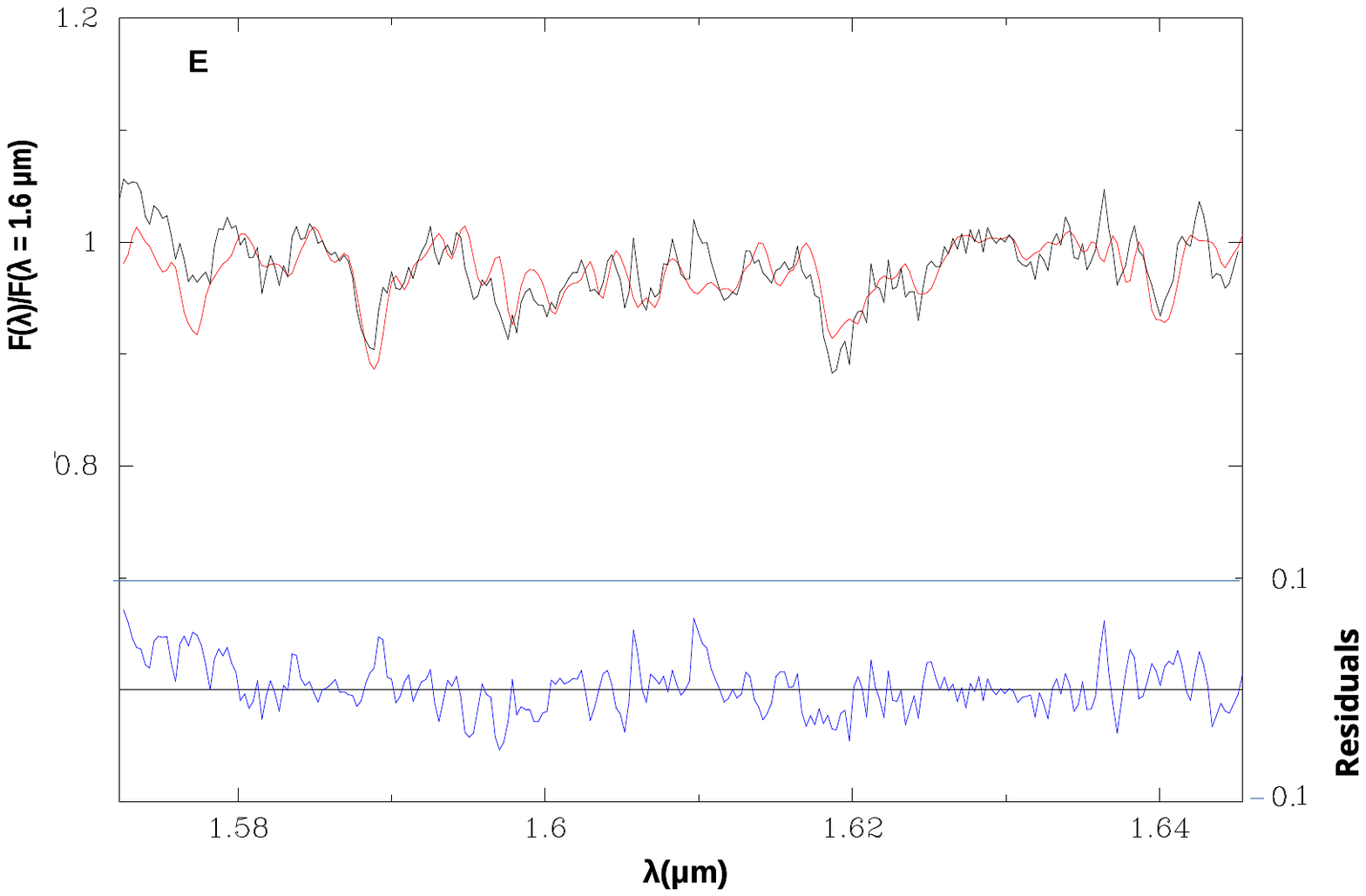}
\includegraphics[width=8.4cm]{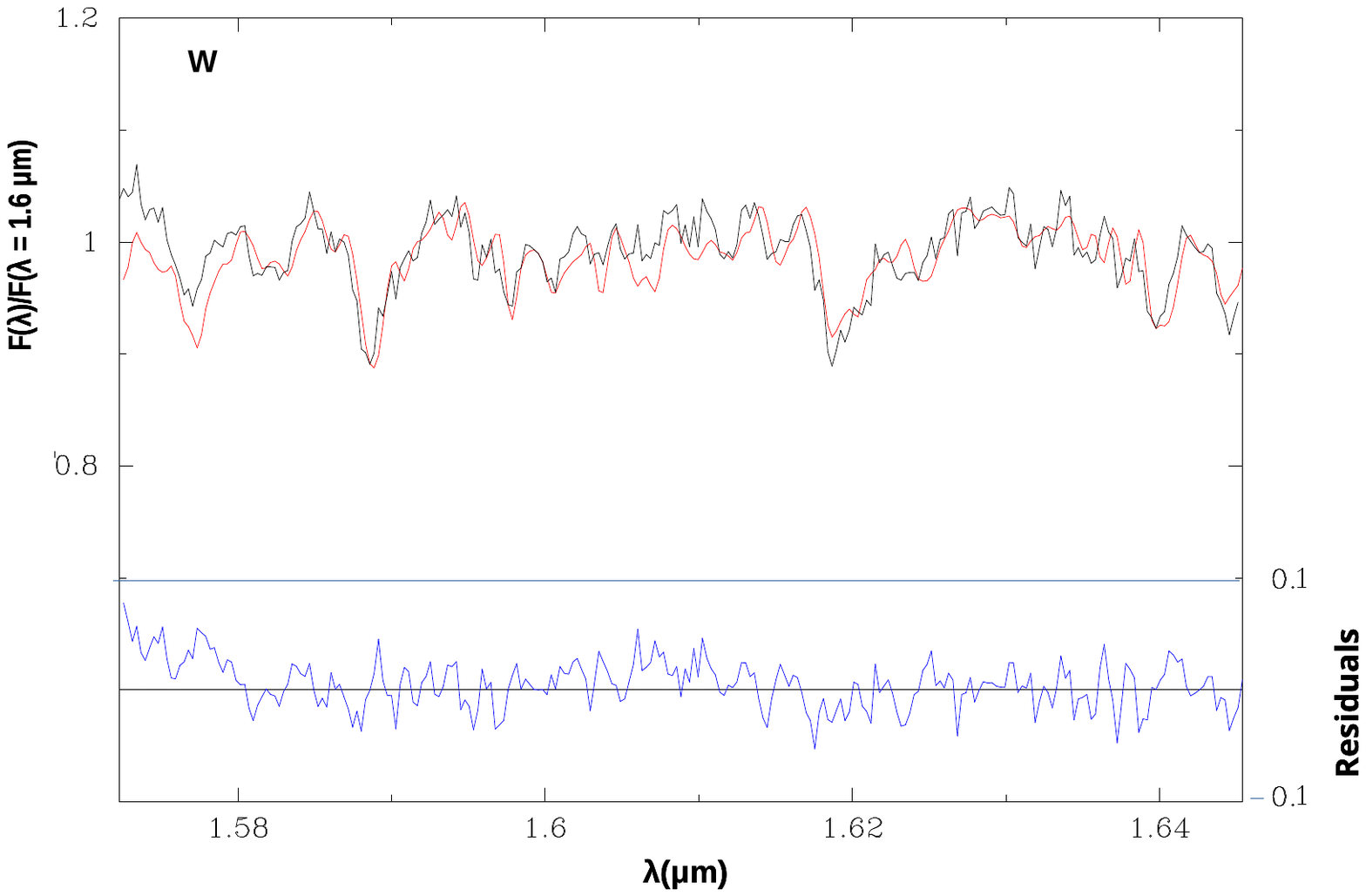}
\caption[]{Stellar population fits to the three regions of MCG-6-30-15
  taken with VLT/ISAAC. From top to bottom: nucleus, east, and west (see 
Fig.~\ref{fig:imMCG} bottom)}
\label{fig:fitMCG3}
\end{figure}

The Seyfert~1 galaxy MCG-6-30-15 has been extensively studied in all wavelength 
bands though primarily in X-rays. The host galaxy is an elongated E/S0.  
Photometric colour maps show a dust lane crossing the galaxy just south of the 
nucleus, roughly parallel to the major axis of the galaxy \citep{Ferruit00}. 
 
From a spectral analysis in the optical, the bulge ($\sim$ 200 - 800 pc) stellar population  has been 
found to be  dominated by old mildly metal-rich stars with a contribution of 
intermediate age stars (around 1 Gyr) that increases farther away from the 
galactic centre, see \cite{Boisson+04}. 

With its high spatial resolution, SINFONI allows one to sample  the inner nuclear regions well.  We analyse the five SINFONI regions (1 to 4 and a pseudo-ring of radius $\sim$1.1 arcsec excluding the very nucleus)  observed by \cite{Raimundo+13}. The ISAAC spectra sample two more regions besides the ones mapped by SINFONI: the nucleus ($<$ 1 arcsec /110 pc radius) and two regions east and west. The regions studied are marked on  the sharp-divided image of the central region 
of  MCG-6-30-15 as shown in Fig.~\ref{fig:imMCG}.

The results of the stellar population fits of the various regions of
MCG-6-30-15 are given in Table~\ref{tab:mcgres}, with the
corresponding spectra and fits shown in Figs.~\ref{fig:fitMCG1} and
\ref{fig:fitMCG2}. We note the excellent agreement between the observed
and synthetic spectra. 

The presence of a non-stellar component is important in all SINFONI 
regions, specially in regions 2 (N) and 4 (S), which are closer to the nucleus, 
and therefore may be more contaminated by the AGN emission. Contrary to the SINFONI 
spectra, the inner 110 pc ISAAC spectrum includes the very nucleus, which then induces a higher contribution of the power-law component (83\%). 

The whole central $<$110  pc radius population is old ($\sim$ 5.0 $\times$10$^{9}$ yrs), dominated by dwarfs and evolved M giant stars. An important contribution of young (10$^{7}$ yrs) supergiant stars (M2I) is also present. Using emission lines to study  the  supernova rate, \cite{Raimundo+17} also deduce the presence of a young $<$10$^{8}$ yrs population. We note that the better spectral resolution of SINFONI, compared to ISAAC, reveals the presence of old super metal-rich dwarf stars as expected in case of successive bursts of star formation. 

As can be seen from the ISAAC data, the outer regions (110-330 pc) are dominated by 
intermediate (F2V, $\sim$ 4.0 $\times$10$^{8}$ yrs) and old stars  with no recent burst of star formation as indeed  expected in lenticular galaxies (see Fig.~\ref{fig:fitMCG3}). At such a large distance from the nucleus, 
the metallicity of the contributing stars is weaker than in the nuclear regions. A $\sim$ 20\% contribution of young star population is present in the nucleus ( $<$ 110 pc).

Analysing in a large IR range (J, H, and K bands) the inner 300~pc, \cite{Riffel+24} found a dominating old population, the youngest contribution they obtain is of intermediate age ($>$10$^{8}$ yrs). This may be due to a lack of spatial resolution.

In conclusion, the nuclear stellar content of the Seyfert 1 MCG--6-30-15 is dominated by a mixture of old and young  populations with the contribution of metal-rich stars proving a history of multiple bursts of star formation. This is in contrast with the overall bulge population inferred from our study in the optical, and points to a population gradient with ongoing star formation in the nucleus.

\subsection{NGC~2110}

\begin{table}
  \caption{Spectral synthesis of NGC~2110. }
\begin{tabular}{rrrrr}
\hline
\hline
             & C(nuc)  & SE      & NW1      & NW2           \\
             & 25 pc & 25-120 pc& 25-120 pc& 120-235 pc    \\
\hline       
$\sigma$ km/s&  100& 100& 100 &  100            \\                                                
\% stell     &  50$\pm$1   &  100$\pm$0 &   100$\pm$0 &  100$\pm$0          \\
\% PL        &  50$\pm$1   &  0   &   0   &  0            \\
\% BB        &  -    &  -   &   -   &  -              \\
T (BB)       &  -    &  -  &  -    & -               \\
E(B-V)       &  0.94$\pm$0.20  & 1.23$\pm$0.17 &  - &  -           \\  
\hline
A0V       &      &         &      &  3$\pm$1    \\
F2V      &      &         &      &   26$\pm$1  \\
rG1V      &      &  24$\pm$1     &  27$\pm$1 &  23$\pm$1  \\
wM3V       &      &  28$\pm$4     &   13$\pm$1   &      \\
F0IV     &      &    12$\pm$1     &   13$\pm$1   &  1$\pm$1   \\ 
F8III     &   26$\pm$2   & 10$\pm$1      &  19$\pm$1     &      \\
G6III     &   22$\pm$3 &      &    &      \\
K5-M0III  &      &  2$\pm$1       &      &   33$\pm$1  \\
M4III     &  52$\pm$1  &   24$\pm$4  &  28$\pm$1  &  14$\pm$1  \\
\hline
\end{tabular}
\\
\label{tab:n2110res}
\end{table}

\begin{figure}
\centering
\includegraphics[angle=0,width=6cm]{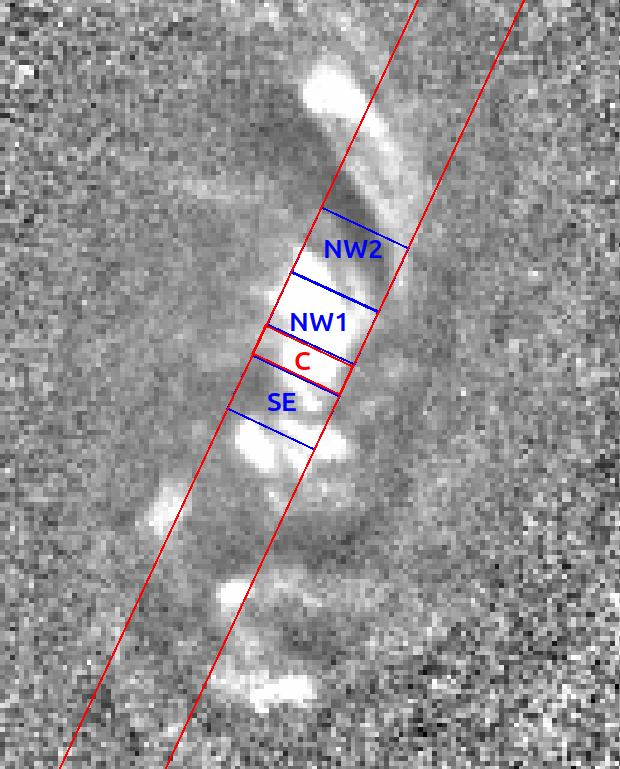}
\caption[]{Sharp-divided IR HST image in the fr680p15 filter of
  NGC~2110 with the ISAAC slit superimposed and the studied regions
  marked.}
\label{fig:imN2110ir}
\end{figure}

\begin{figure}
\centering
\includegraphics[width=8.4cm]{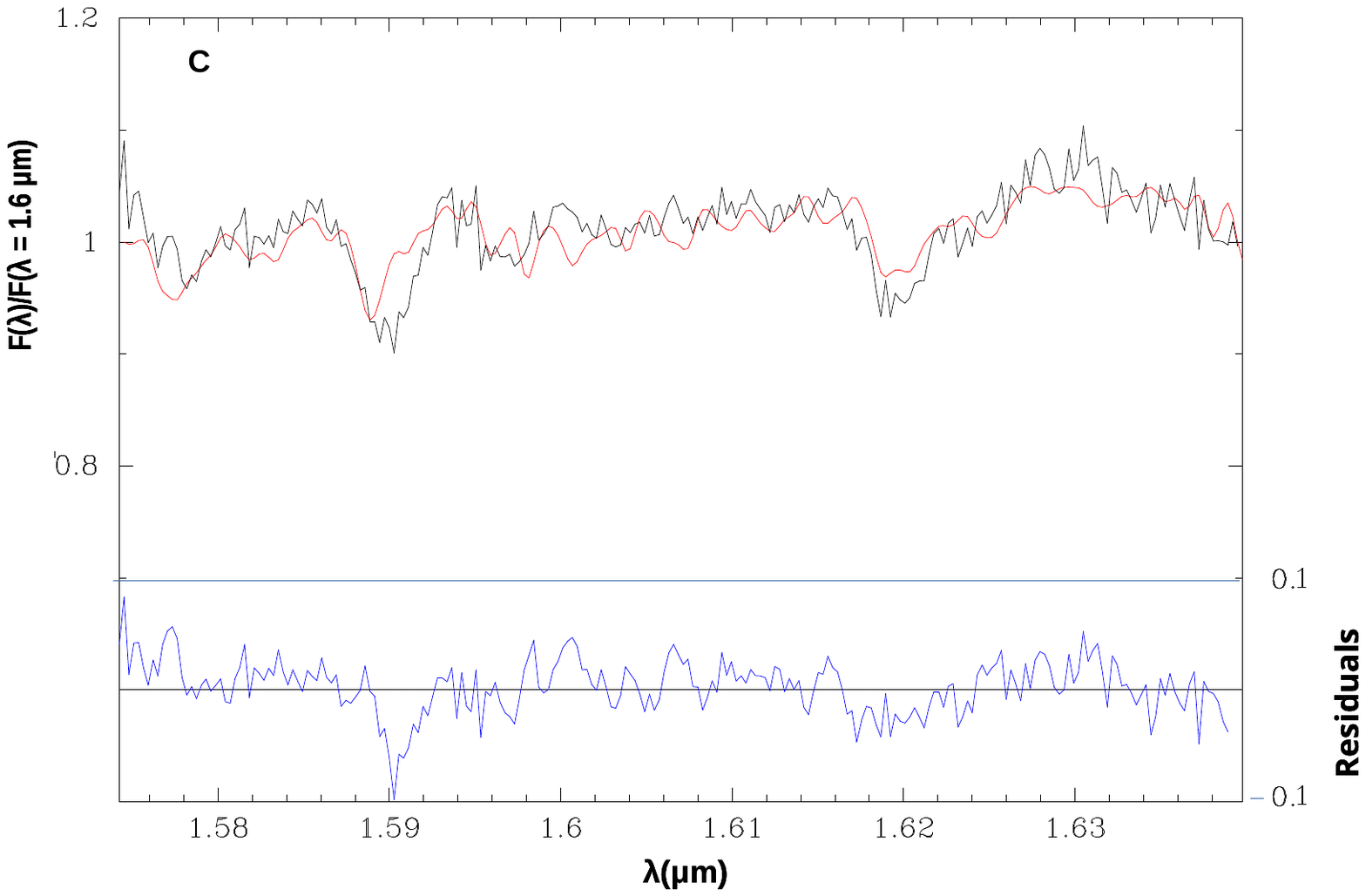}
\includegraphics[width=8.4cm]{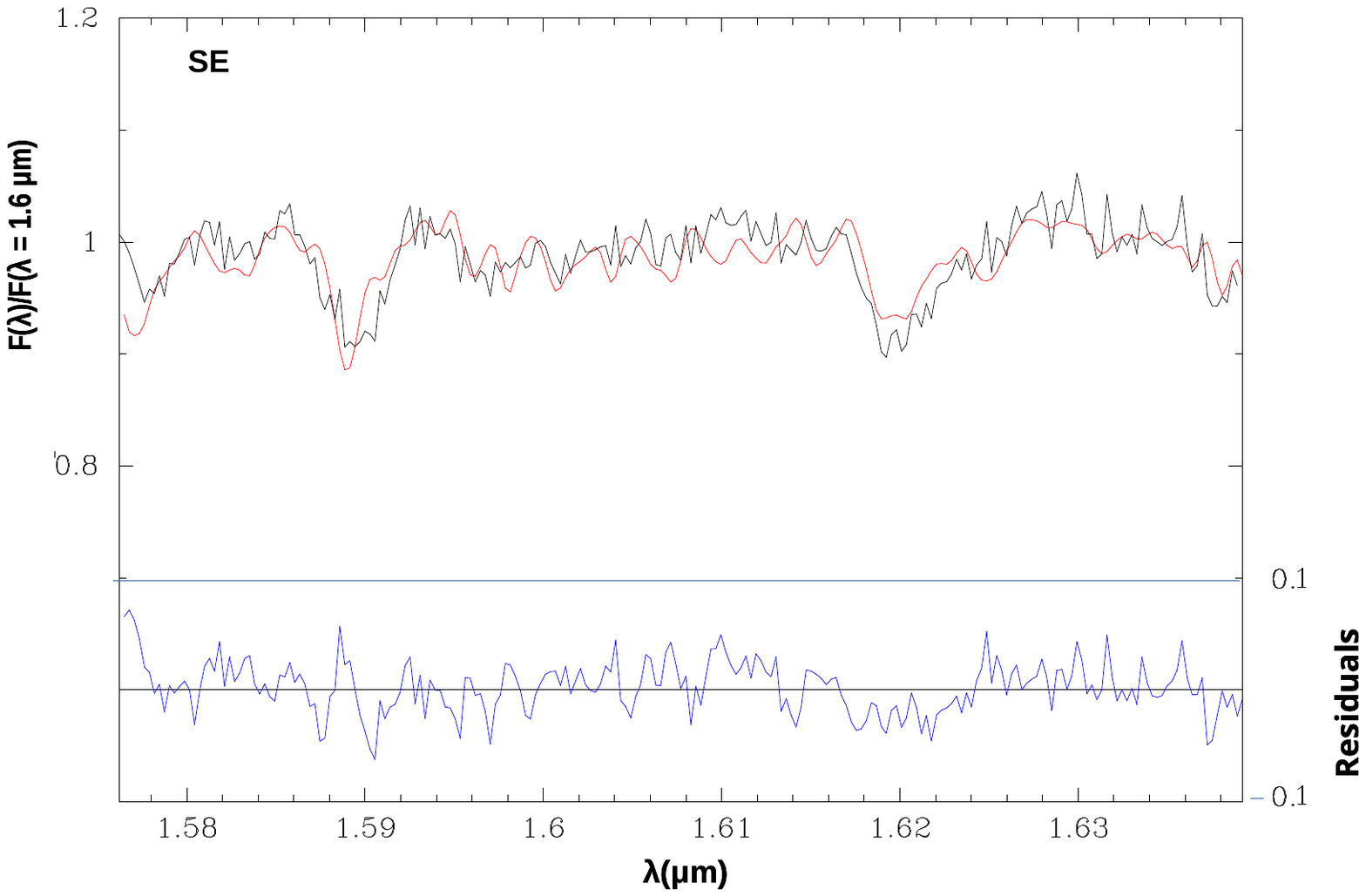}
\includegraphics[width=8.4cm]{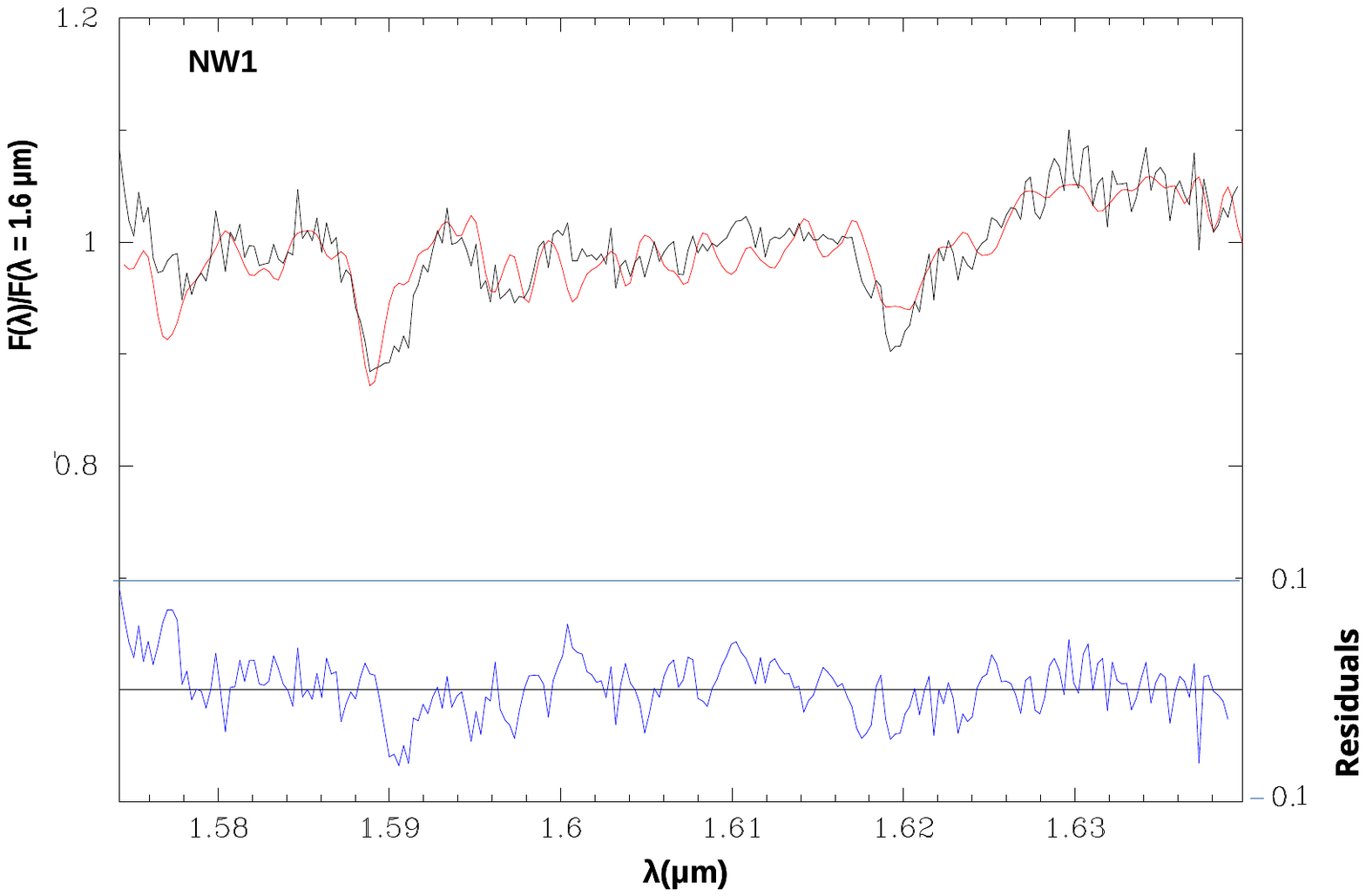}
\includegraphics[width=8.4cm]{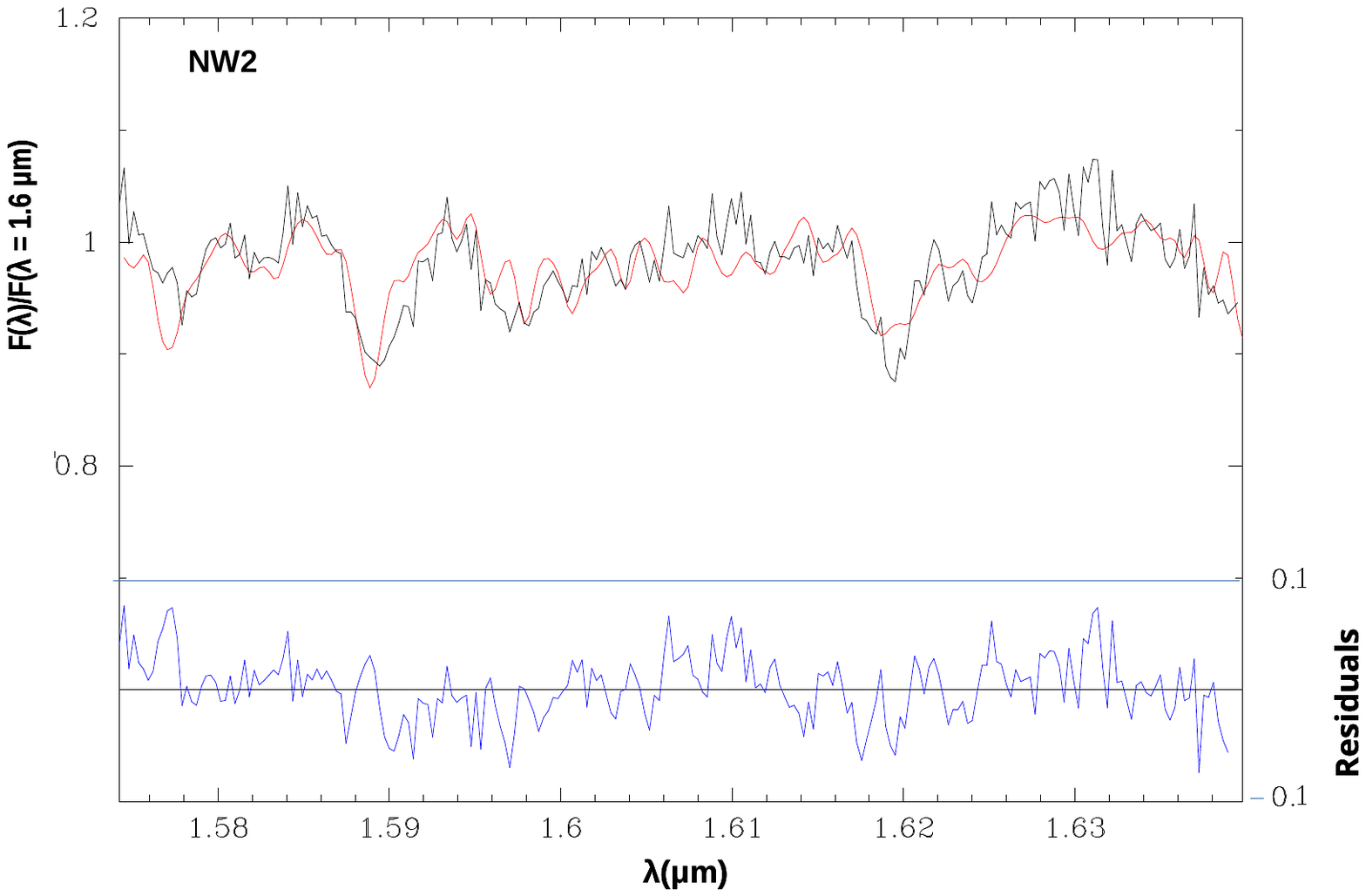}
\caption[]{Stellar population fits to the four regions of
  NGC~2110. From top to bottom: the nucleus, the SE, and the 
  two NW regions (see Fig.~\ref{fig:imN2110ir})}.
\label{fig:fitN2110}
\end{figure}

NGC~2110 is a weakly barred S0 galaxy harbouring a Seyfert~2 nucleus, also 
classified as a FRI radio galaxy \citep{Beckmann2010}. High spatial resolution HST continuum images show a dusty circumnuclear disk with a spiral  pattern \citep[{see also Fig. 5}]{Evans+06, SchnorrMuller+14}. 

The sharp-divided image 
of NGC~2110 is shown in Fig.~\ref{fig:imN2110ir}
with the studied regions superimposed. We extracted the spectra of four regions: a nuclear region 50x150~pc$^2$ (C), two regions on
each side of the nucleus extending from 25 to 120 pc (SE and NW1) and 
a second adjacent region NW up to 235 pc from the nucleus (NW2).

The stellar population in the nucleus of NGC 2110 is dominated by giant evolved stars.  
In the ring surrounding the nucleus, up to 120 pc, an old population is still dominating, but includes  partly metal-rich and partly weak metallicity dwarf stars. The farthest NW region, beyond 120 pc of the centre, shows a large contribution of intermediate stars indicative of an ancient ($\sim$ 3.0 $\times$10$^{8}$ yrs) star formation burst in the past history of the galaxy. This population gradient is in line with the results of a study of the inner 3x3~arcsec$^2$ (450x450 pc$^2$)  by \cite{Diniz+19} through near IR J and K bands integral field spectroscopy.
 
We do not detect the strong signature of young stars in the central 0.6~arcsec as claimed from optical integral field spectroscopy by \cite{Dahmer+22}. Neither do \cite{Burtscher+21} in a similar study, who point to a largely dominating old stellar population. In the near IR,  \cite{Riffel+24} found a stellar population of mean age $\sim$ 5~Gyr in  a region that includes our three innermost regions, in  global agreement with us. And though we detect sub-solar metallicity dwarfs as in \cite{Dahmer+22}, we also highlight a contribution of metal-rich stars in this same region. 

The synthetic spectra are displayed in Fig.~\ref{fig:fitN2110}. It is worth noticing that, in the four regions analysed, residuals are clearly marked around 1.59~$\mu$m and 1.62~$\mu$m, where one expects a combination of several molecular features (SiI, FeI, OH, CO, CaI). Such line strengths and ratios vary with both stellar temperature and gravity, implying that  ultra-cool dwarfs are missing in the data base we use. NGC~2110 is the only galaxy in the small sample studied where such a lack appears, implying a peculiar stellar population for this object. 

In conclusion, the stars that contribute the most to the stellar population in the inner 235 pc are old and cold, indicating the absence of star formation since $\sim$ 6.0 $\times$10$^{9}$ yrs, although a younger population is present in the outer regions. The presence of a metal-rich star contribution shows that in the past there have been many star formation episodes. This was suggested in the optical study \citep{Boisson+00} where, in the whole bulge up to 800 pc, an old but metal-rich stellar population is the proof of multiple star formation episodes in the past.

\subsection{NGC~2992}

\begin{table}
  \caption{Parameters of the spectral synthesis for NGC~2992.}
\begin{tabular}{rrrrr}
\hline
\hline
         &  C(nuc)   &  SW  &  NE \\
         &  110 pc &  110-330 pc &  110-330 pc\\           
\hline
$\sigma$  km/s    &   95  &  80  &  90  \\
\% stell &  38$\pm$1  &   64$\pm$2    &   98$\pm$0           \\
\% PL    &  62$\pm$1   &  0     &   0            \\
\% BB    & -   &  36$\pm$2    &   2$\pm$0              \\
T (BB)   &  -    &   970$\pm$23 &   200$\pm$0              \\
E(B-V)   &  0.70$\pm$0.07   &  -  &  -             \\
\hline
F2V      &        &    &  19$\pm$1   \\
rG1V      &    14$\pm$5  &  36$\pm$2  & 13$\pm$1      \\
K4V       &     5$\pm$2  &    &       \\
wM3V       &      &    &   14$\pm$1    \\
F0IV      &       &  3$\pm$1  &        \\ 
F8III     &       &    &    12$\pm$1  \\
M4III     &   81$\pm$1  &  34$\pm$1  &  42$\pm$0    \\
M2I       &       &  27$\pm$0  &       \\
\hline
\end{tabular}
\label{tab:n2992res}
\end{table}

\begin{figure}
\centering
\includegraphics[angle=0,width=7cm]{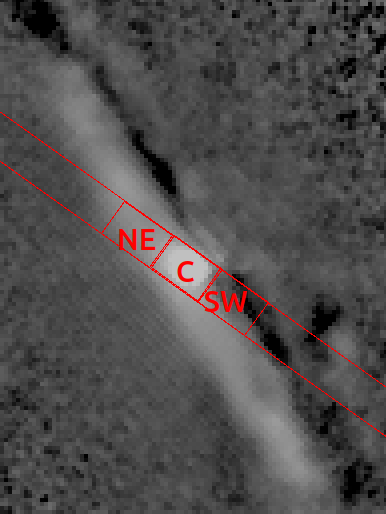}
\caption[]{Sharp-divided IR HST image in the F196N filter of
  NGC~2992 with the ISAAC slit superimposed and the three regions
  indicated: the nucleus (C), the north-east (NE), and south-west (SW) regions.  
  }
\label{fig:imN2992}
\end{figure}

\begin{figure}
\centering
\includegraphics[width=8.4cm]{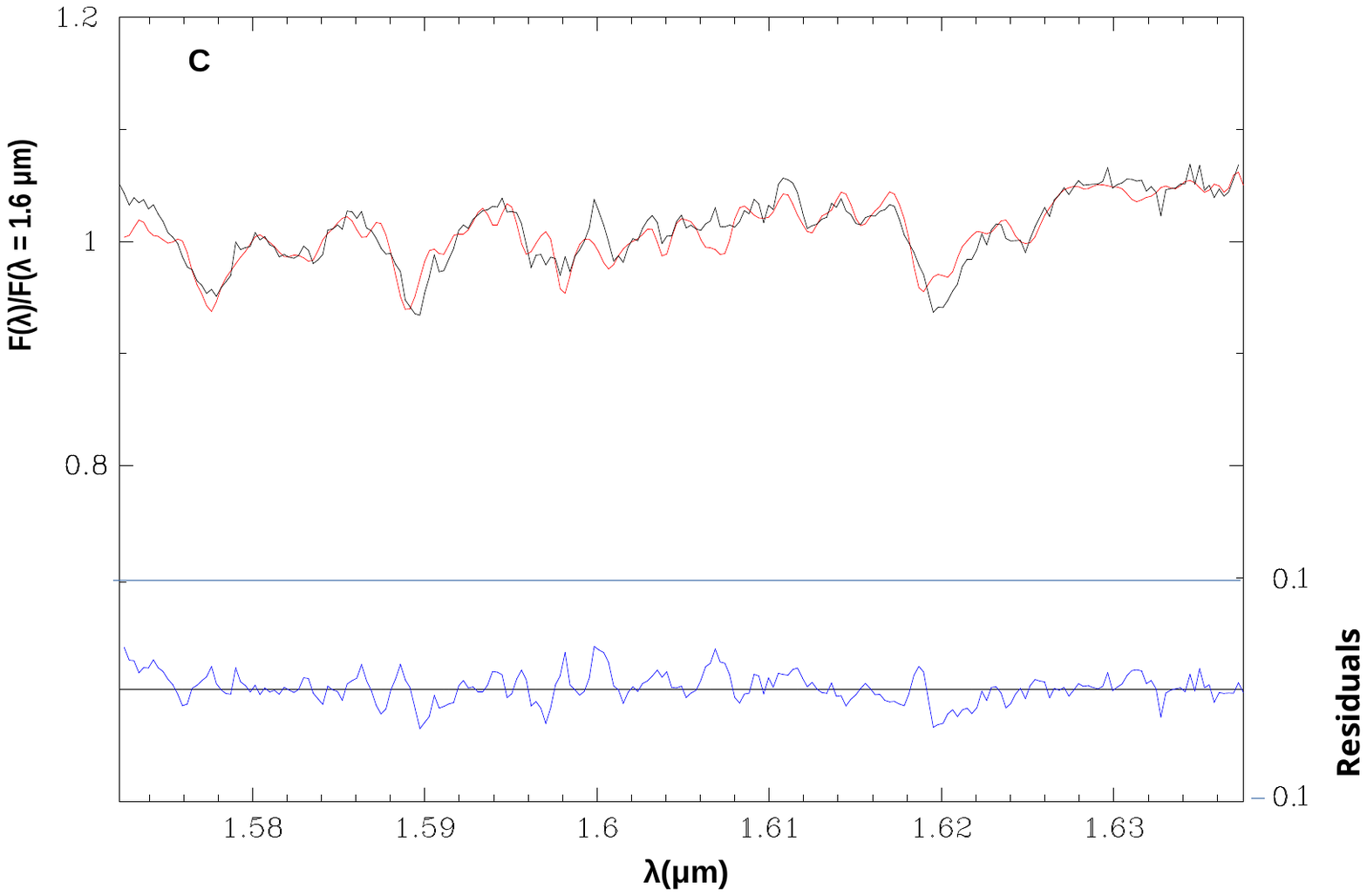}
\includegraphics[width=8.4cm]{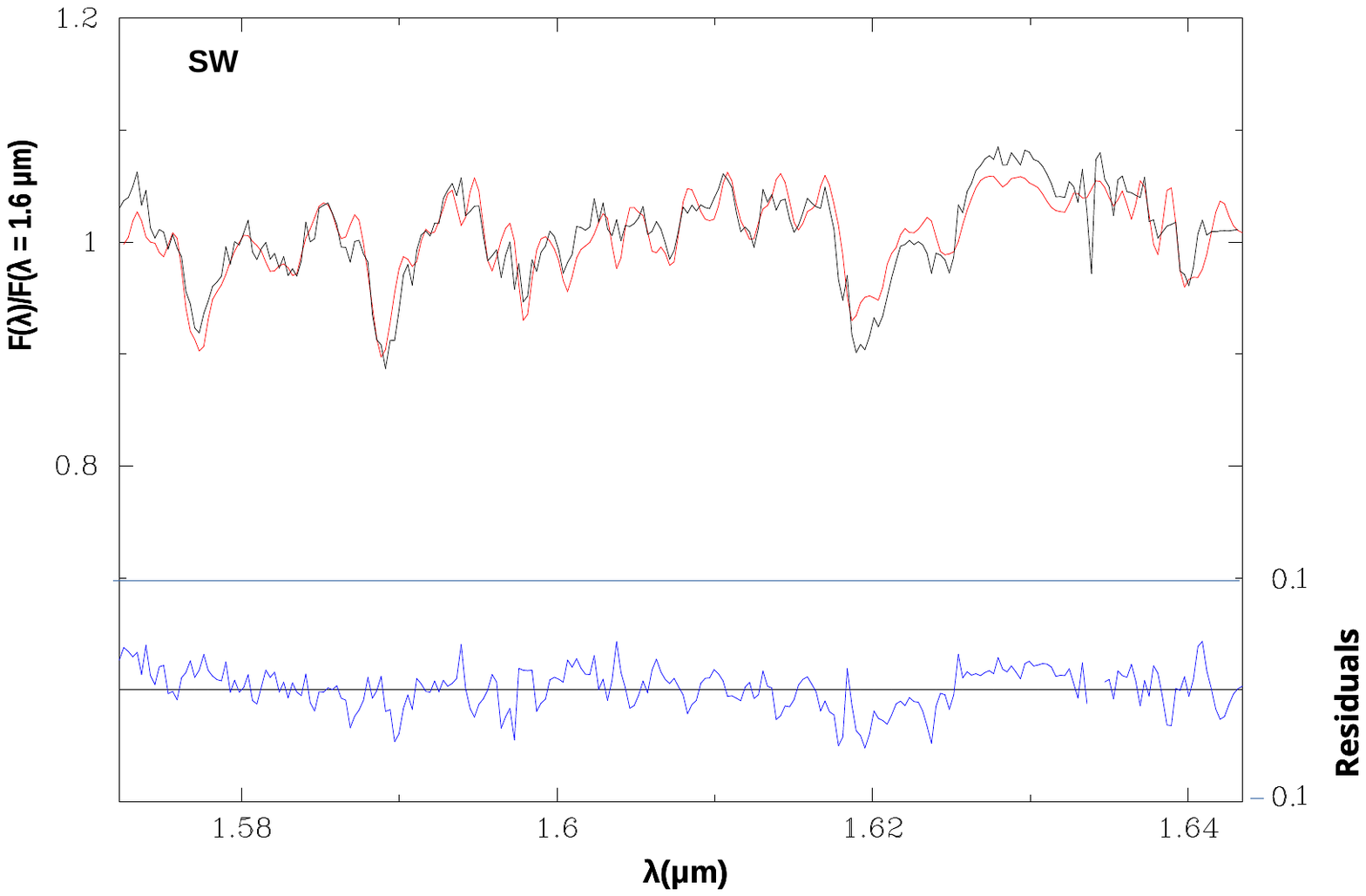}
\includegraphics[width=8.4cm]{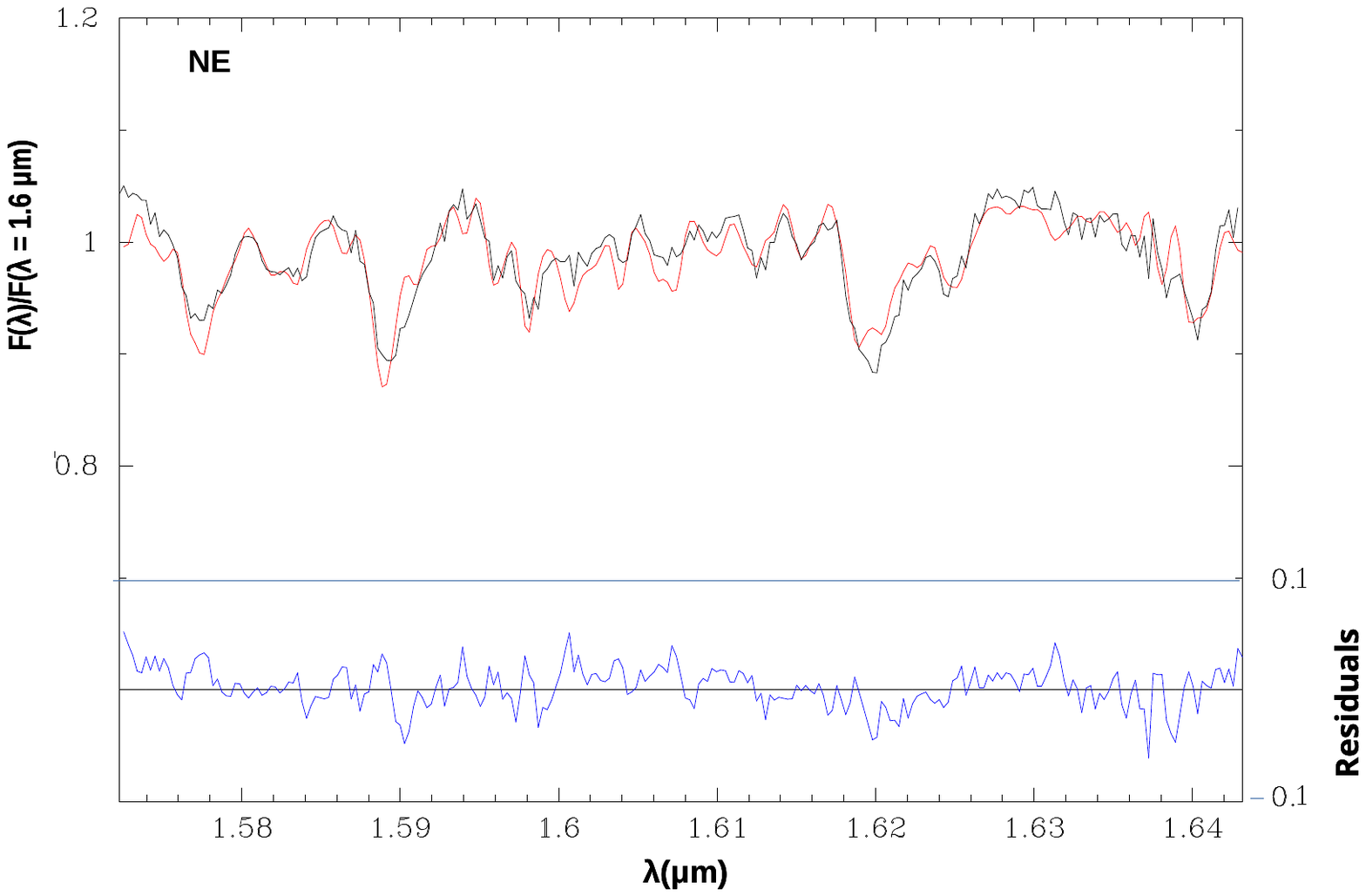}
\caption[]{Stellar population fits to the three regions of
  NGC~2992. From top to bottom: nucleus, SW, and NE (see
  Fig.~\ref{fig:imN2992}).}
\label{fig:fitN2992}
\end{figure}

NGC 2992 is a Sa galaxy seen nearly edge-on,  interacting with its neighbour NGC 2993. 
A prominent dust lane extending along the major axis crosses the nucleus of the 
galaxy \citep[see also Fig. 7]{Ward+80}. NGC 2992 is classified as a Seyfert 1.9 in the 
optical, although it has changed its type between Seyfert 1.5 and 2 in the past \citep{Trippe+08}. A strong broad OI$\lambda$8446~\AA\  emission line, which is a 
clear signature of Seyfert 1s not seen in Seyfert 2s, was indeed visible when this highly variable galaxy was in a high state \citep{Boisson+04}. Thus, in any statistical study
the properties of this galaxy may be compared to those of Seyfert 1s and 
not of Seyfert 2s.  This galaxy exhibits a biconical galactic-scale outflow, which emerges almost perpendicularly from the plane of the galaxy \citep{Marquez+98, Friedrich+10}.
The sharp-divided image of NGC~2992 is shown in Fig.~\ref{fig:imN2992} with the studied regions superimposed. 

In a detailed IR adaptive optics study, \cite{Chapman2000} found a radial distribution of the CO index in  the inner 3 arcsec ($\sim$ 450 pc), suggesting a  population gradient within the core, with the stellar population  at the very centre being older than that in the surrounding regions. Stellar population modelling of OASIS integral field spectroscopy by \cite{Stoklasova2009} revealed a contribution of young stars close to the dust lane, younger than in the rest of their field of view, with mostly old stars in the eastern part.   From optical GMOS observations, \cite{Guolo-Pereira2021+} find a mainly old metal-rich stellar population, with a noticeable contribution of young metal-poor stars in the inner 1.1 kpc of NGC~2992. \cite{Burtscher+21} from VLT/X-shooter spectra in an aperture of 150 parsec radius infer a dominating rather old reddened stellar population ($>10^{9}$ yrs). Another study by \cite{Dahmer+22}  with MUSE integral field spectroscopy of the 270 central parsecs, points to a mainly young, metal-rich, and highly reddened stellar population to the north–west, with the rest of their field of view (FoV) having an old, metal-poor, and less reddened stellar population. 
However, a study in the IR by \cite{Riffel+24}, with SpeX attached to the IRTF, sampling the same dimensions, infers a largely dominating old population ($\sim$ 8 Gyr). 

In Table~\ref{tab:n2992res} the results of the stellar population fits for the three regions of NGC~2992 we studied are given, with the corresponding spectra and fits shown  in Fig.~\ref{fig:fitN2992}.
The AGN non stellar contribution reaches about 60\% in the nucleus. We find an evolved (M giant stars, $4\times 10^9$~yrs) stellar population of about solar metallicity, in a high extinction environment, with no hint of a young stellar population contribution. The stellar population is younger in the regions surrounding the nucleus, with not more than the foreground galactic extinction.  An important star formation difference is visible  between the SW and NE regions: presence of an important metal-rich component and recent ($\sim 10^{7}$ yrs) starburst in the SW region, the closest to the dust lane, and an older burst of star formation inducing  an intermediate age stellar population in the NE region. 

The hot dust temperature inferred by the analysis of the SW region is not expected. 
A test constraining the dust temperature to 500~K induces an increase of the computed extinction up to the nuclear one, a plausible result as this region is the closest to the dust lane. The global stellar population remains the same, pointing to some degeneracy between extinction and featureless continuum from  dust emission.
In our study of the 600 central parsecs, taking advantage of the H-band, we can emphasise a clear age and possible metallicity gradient, besides a high star formation in the SW region.

\subsection{NGC~3185}

\begin{table}[h!]
  \caption{Parameters of the spectral synthesis for NGC~3185. }
\begin{center}
\begin{tabular}{rrr}
\hline
\hline
         & C (nuc)       & R (ring)     \\
	 & 60 pc    &  60-175 pc\\       
\hline
$\sigma$  km/s &  75   &  80    \\
\% stell &  100$\pm$0  & 100$\pm$0     \\
\% PL    &  0    & 0      \\
\% BB    &  -    & -      \\
T (BB)   &  -    & -      \\
E(B-V)   &  - & -   \\
\hline
 F2V      &       & 10$\pm$0     \\
 rG1V     &  14$\pm$1    & 28$\pm$0     \\
 K4V      &   11$\pm$1  &       \\
F0IV      & 18$\pm$1    &  14$\pm$0   \\ 
F8III     &   4$\pm$1   &       \\
M4III     & 53$\pm$1    & 48$\pm$0    \\ 
\hline
\\
\end{tabular}
\end{center} 
\label{tab:n3185res}
\end{table}

\begin{figure}
\centering
\includegraphics[width=7cm]{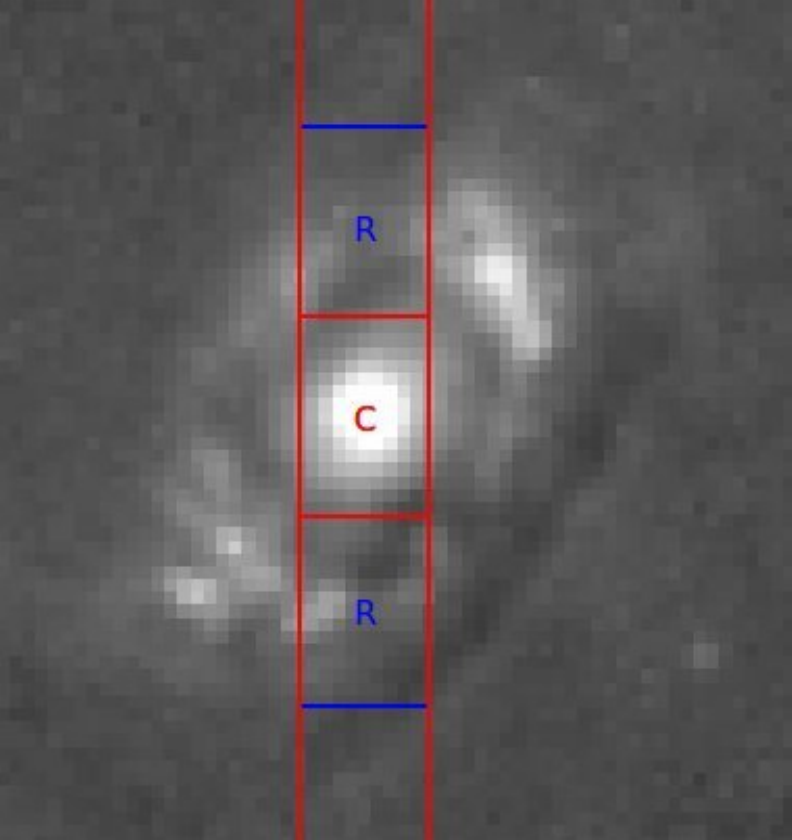}
\caption[]{Sharp-divided 
HST image in the F814W filter of
  NGC~3185 with the ISAAC slit superimposed. The nuclear region is
  indicated by C, and the ring corresponds to the sum of the two blue
  regions.}
\label{fig:imN3185}
\end{figure}

\begin{figure}
\centering
\includegraphics[width=8.4cm]{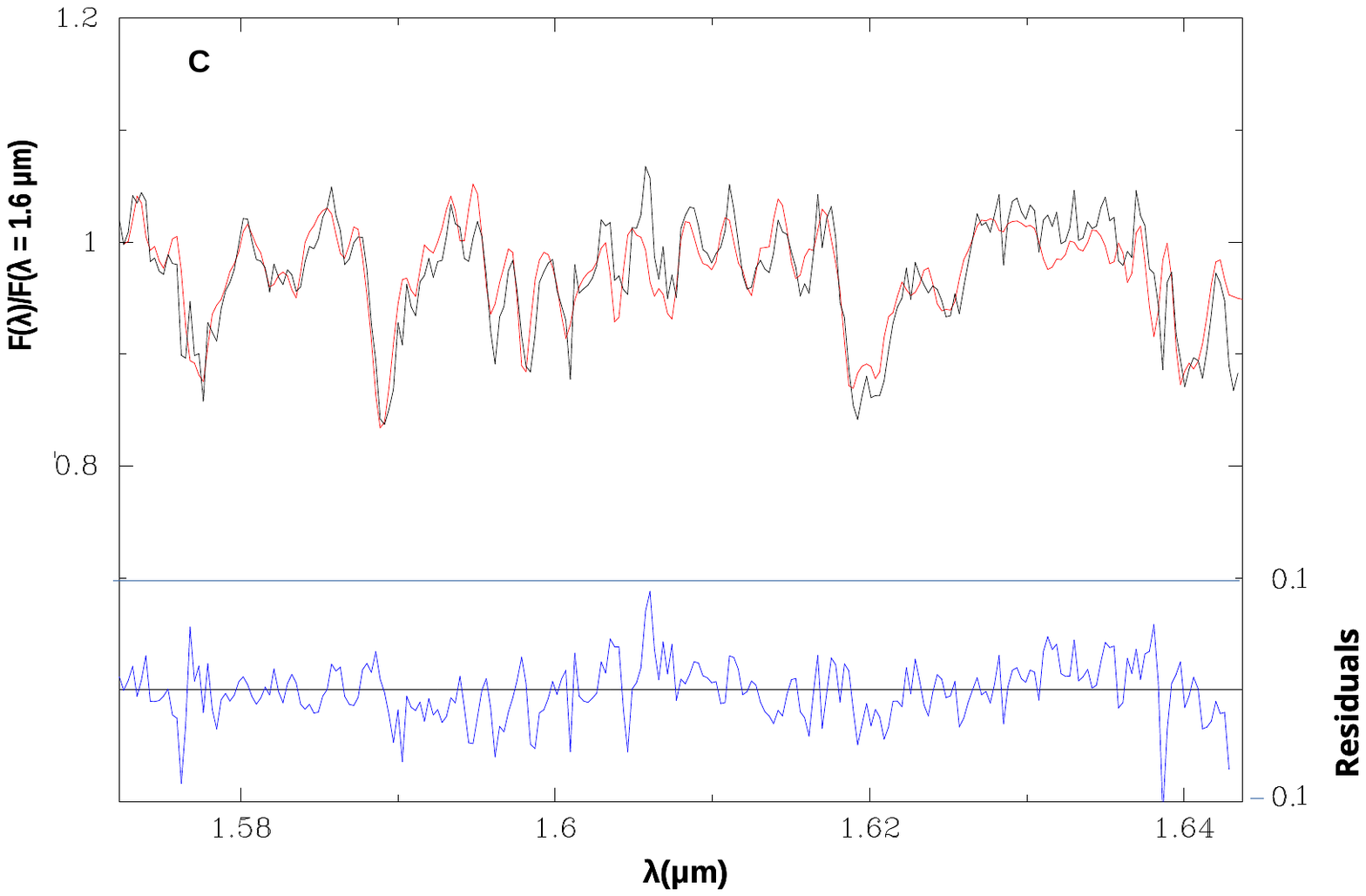}
\includegraphics[width=8.4cm]{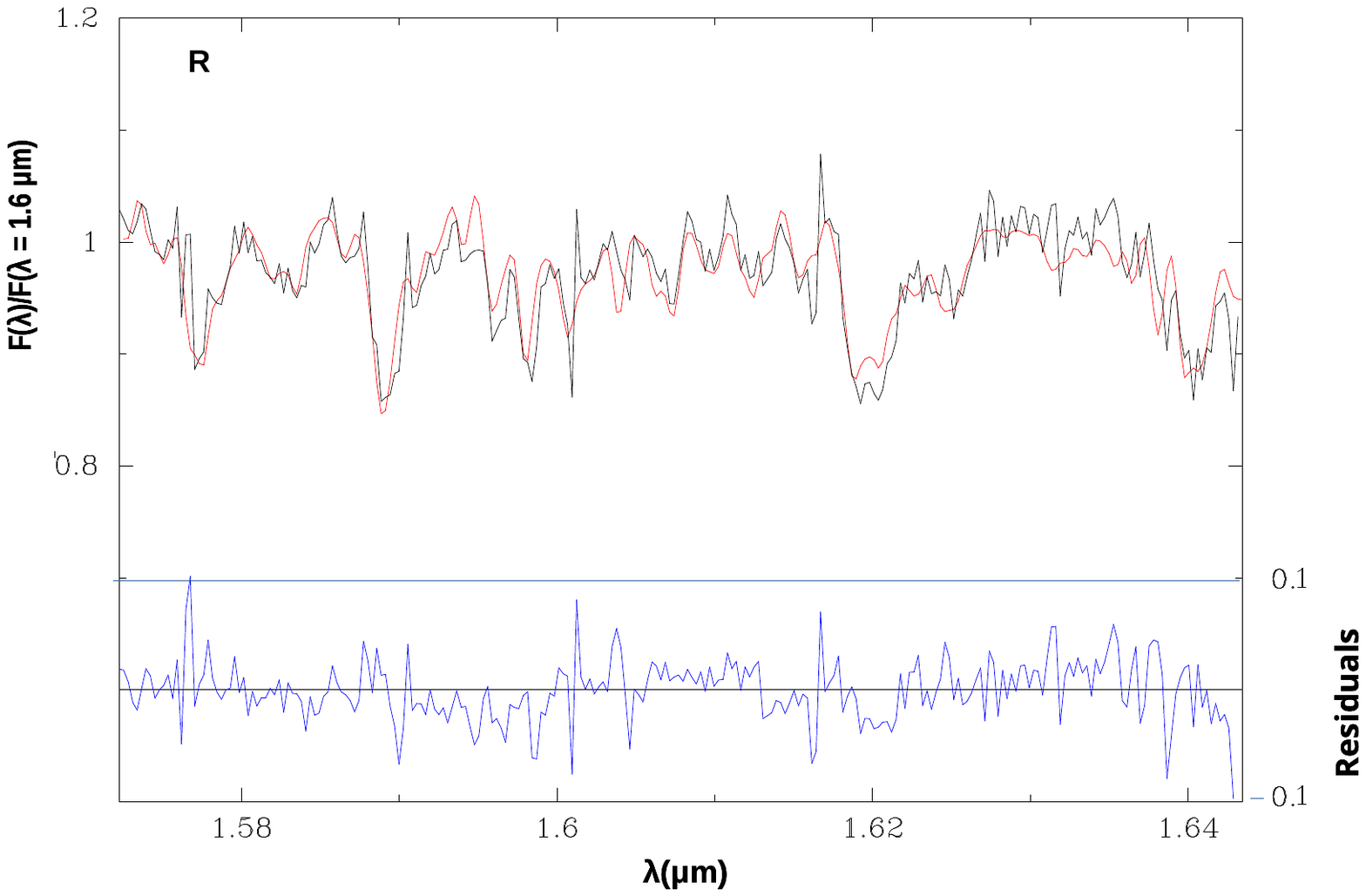}
\caption[]{Stellar population fits to the two regions of
  NGC~3185. From top to bottom: nucleus and ring (see
  Fig.~\ref{fig:imN3185}).}
\label{fig:fitN3185}
\end{figure}

NGC~3185 is a strongly barred SBa galaxy located in a compact Hickson group dominated by NGC~3190.  This galaxy classified as Seyfert~2 by \cite{Filippenko85}, is also considered as a transition object with both Seyfert and star forming characteristics \citep[e.g. ][]{Goncalves_1999}. A LINER classification was also proposed by \cite{Carrillo+99}.
Circumgalactic rings of star forming material are observed in both UV and IR images on a large scale \citep{Lanz+13}.

From optical studies, NGC~3185 has been shown to have a very old population in the central 1.56 kpc 
\citep{James+16}.   To our knowledge, 
there are no other published results on the stellar content of the nucleus of 
this galaxy.  

The sharp-divided image of the galaxy is shown in Fig.~\ref{fig:imN3185} with 
the studied regions superimposed. The spectra of the two regions on either side 
of the nucleus are very similar and thus combined to represent the spectrum of a 
ring-like feature at about 120~pc from the nucleus. The stellar population derived for the nucleus and ring  are given in Table~\ref{tab:n3185res}, with the corresponding spectra  shown in Fig.~\ref{fig:fitN3185}.  

Contrary to what is expected for a genuine Seyfert nucleus, no dilution by a 
power law component is needed to account for the nuclear spectrum.  Within the 
central 360 pc of NGC~3185, the stellar population is old, partly metal-rich, though 
intermediate-age dwarf F stars in the ring indicate some stellar activity in the 
past. No extinction is found in excess of the galactic value. This points towards 
a currently quiet stellar environment and low activity in the central engine.

\subsection{NGC~3783}

\begin{table}
  \caption{Parameters of  the spectral synthesis for NGC~3783.  }
\begin{center}
\begin{tabular}{rr}
\hline
\hline
        &  R (ring)  \\
        & 140-470 pc\\
\hline
$\sigma$  km/s&  60\\
\% stell & 100$\pm$1 \\
\% PL    &  0    \\
\% BB   &  -  \\
T (BB)  & -  \\
E(B-V)  &  0.01$\pm$0 \\
\hline
F2V    &  33$\pm$1\\
M3V    & 42$\pm$1  \\
F8III  & 5$\pm$0     \\
M4III     & 18$\pm$2  \\ 
M2I     &   2$\pm$1  \\
\hline
\end{tabular}
\\
\end{center}
\label{tab:n3783res}
\end{table}

\begin{figure}
\centering
\includegraphics[angle=0,width=7cm]{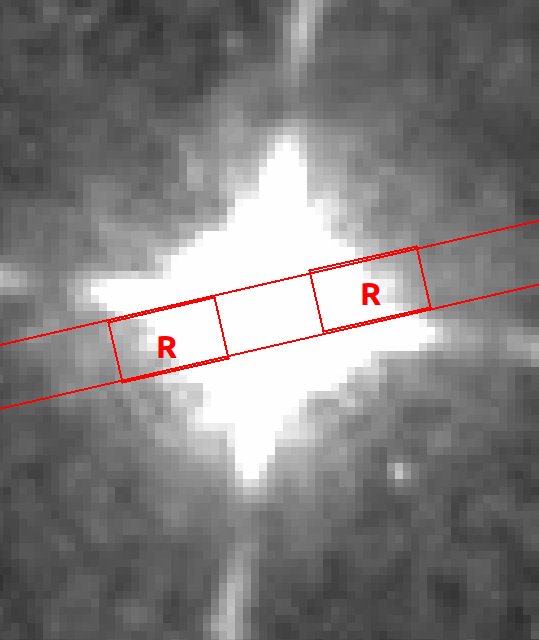}
\caption[]{ HST image in the F160W filter of NGC~3783 with the ISAAC
  slit superimposed. The ring corresponds to the sum of the two 
  regions labelled R.}
\label{fig:imN3783}
\end{figure}

\begin{figure}
\centering
\includegraphics[width=8.4cm]{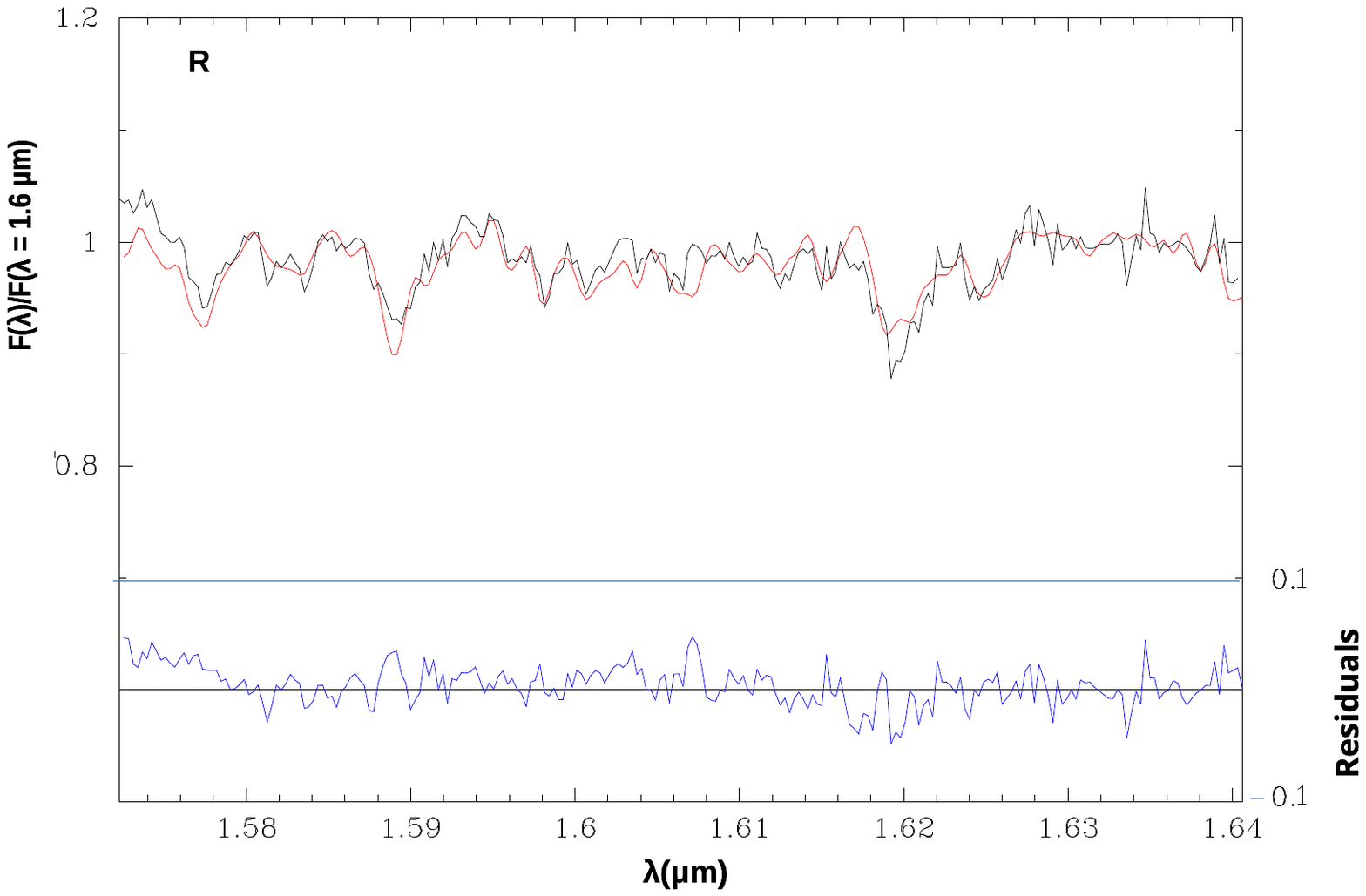}
\caption[]{Stellar population fit of the ring around the central region of NGC~3783 (see Fig.~\ref{fig:imN3783}).}
\label{fig:fitN3783}
\end{figure}

NGC~3783 is a nearly face-on SBa galaxy with an unobscured very bright and 
variable Seyfert  1 nucleus \cite[e.g. ][]{Bentz+09}. The 
sharp-divided image of  NGC~3783 is shown in Fig.~\ref{fig:imN3783} with the 
studied region superimposed; it clearly shows the high contamination from the Sy1 nucleus in this galaxy.

Though NGC~3783 is extensively studied at all frequencies, the strong non-stellar luminosity in the centre makes difficult any quantitative study of the stellar population in the nucleus.  \cite{Davies+07}, from near-IR data, and \cite{Esquej+14},  based on PAH measurements, could only provide an indication of the last burst around $\sim70$~Myr. 
Based on the synthesis of a SOAR near-IR spectrum, \cite{Riffel+24} also inferred a hint of a young stellar population on top of the large contribution of an intermediate stellar population, leading to a mean age of $\sim$~2~Gyr.

We made no attempt to recover the nuclear  stellar population from our data in the central 140 pc,  since the very strong AGN emission overwhelms the  nuclear flux. 
As NGC~3783 is viewed almost face-on with a faint symmetrical bulge structure, a hereafter called ring-like spectrum is built by adding the very similar spectra extracted on either  side of the nucleus \citep{Boisson+02}. The inferred stellar population is given in Table~\ref{tab:n3783res}, and the
corresponding spectrum and fit are shown in Fig.~\ref{fig:fitN3783}.  

Though the stellar population is dominated by about 65\% of old and evolved stars ($>$ 4.0 $\times$ 10$^{9}$yrs), a strong 
contribution of intermediate age stars (33\%) is present as well as a hint of a recent star formation event. We may see the trace of a young population in this region surrounding the nucleus of NGC~3783.

\subsection{NGC~6221}

\begin{table}
  \caption{Parameters of the spectral synthesis for NGC~6221.  }
\begin{tabular}{rrrr}
\hline
\hline
         &  C (nuc)     & N1  & N2      \\
         & 65 pc     &65-170 pc&170-355 pc\\
\hline
$\sigma$  km/s&  60   & 70   & 70   \\
\% stell &  42$\pm$1   & 100$\pm$0   & 100$\pm$0 \\
\% PL    &  58$\pm$1   &  0    &  0   \\
\% BB    &  -    & -     & -    \\
T (BB)   &  -    & -     & -  \\
E(B-V)   &  -  & 0.65$\pm$0.05  & 0.97$\pm$0.04  \\
\hline
A0V       &       &      &  1$\pm$0    \\
F2V       &       &  5$\pm$1   & 17$\pm$1    \\
rG1V      &       & 12$\pm$1   & 7$\pm$2     \\
K4V       &   23$\pm$1   &      &      \\
rM1V     &       &      &   1$\pm$0   \\
wM1.5V     &   3$\pm$2    &    &    \\
wM3V       &   11$\pm$2   & 20$\pm$1     &  17$\pm$2   \\
F8III     &       &  18$\pm$1  &    3$\pm$0  \\
K3III     &       &      & 2$\pm$2   \\
K5-M0III  &       &      &  9$\pm$1   \\
M4III     & 20$\pm$1    & 45$\pm$0   &  34$\pm$1  \\ 
rK2I     &       &      &  7$\pm$1    \\
rK4I      &      &      &  2$\pm$0    \\
M2I       & 43$\pm$1    &      &      \\
\hline
\end{tabular}
\\
\label{tab:n6221res}
\end{table}

\begin{figure}
\centering
\includegraphics[angle=0,width=7cm]{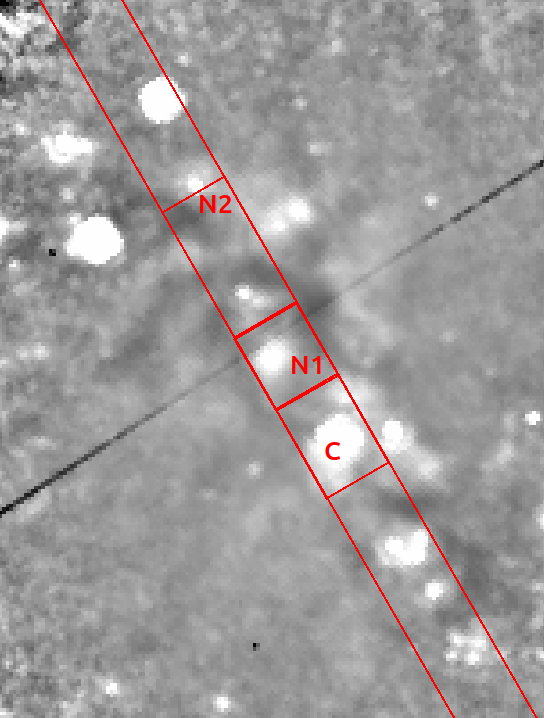}
\caption[]{Sharp-divided IR HST image in the F160N filter of
  NGC~6221 with the ISAAC slit superimposed. The three regions are
the nucleus (C) and the two regions to the north, N1 and N2. }
\label{fig:imN6221}
\end{figure}

\begin{figure}
\centering
\includegraphics[width=8.4cm]{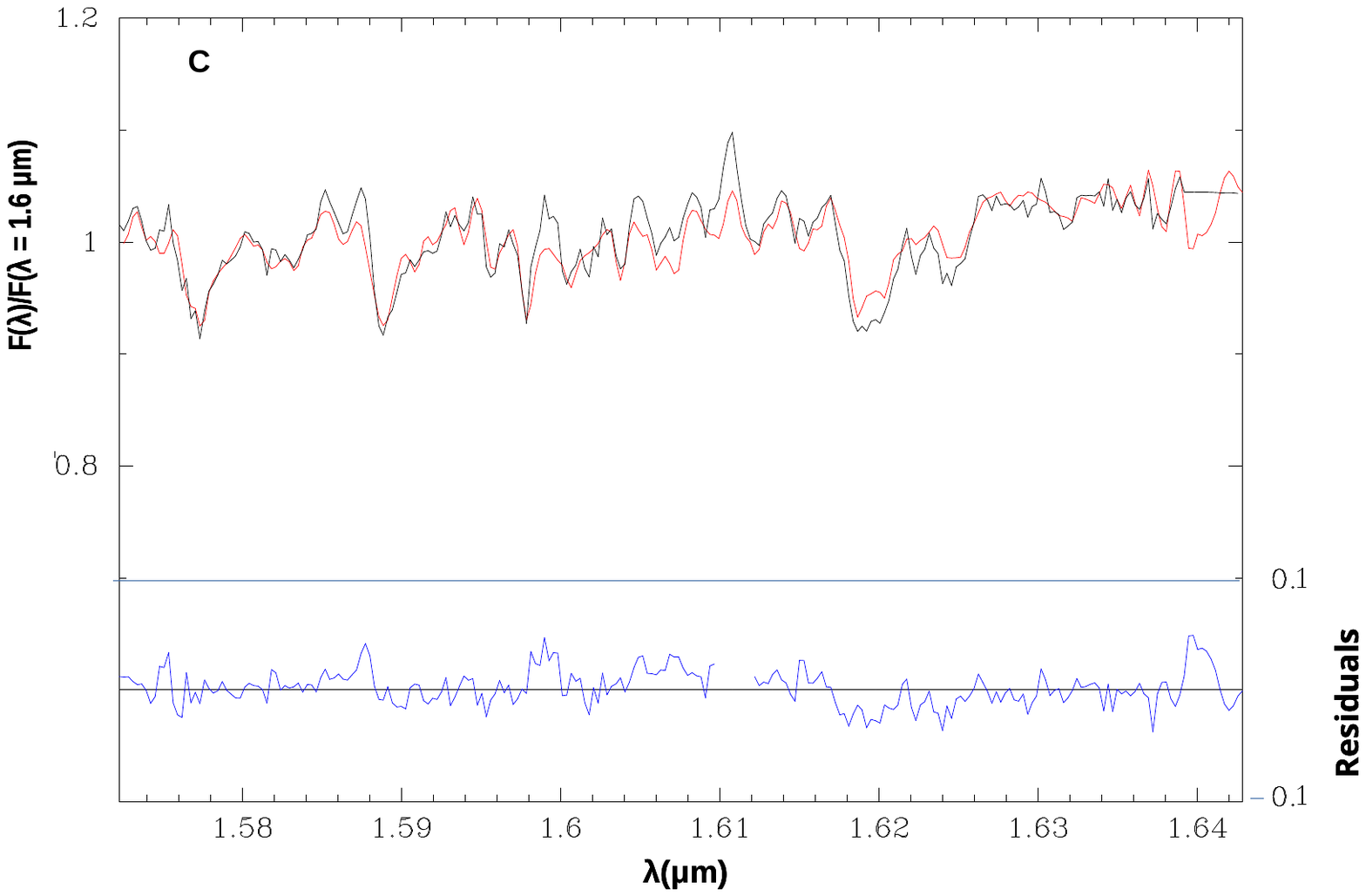}
\includegraphics[width=8.4cm]{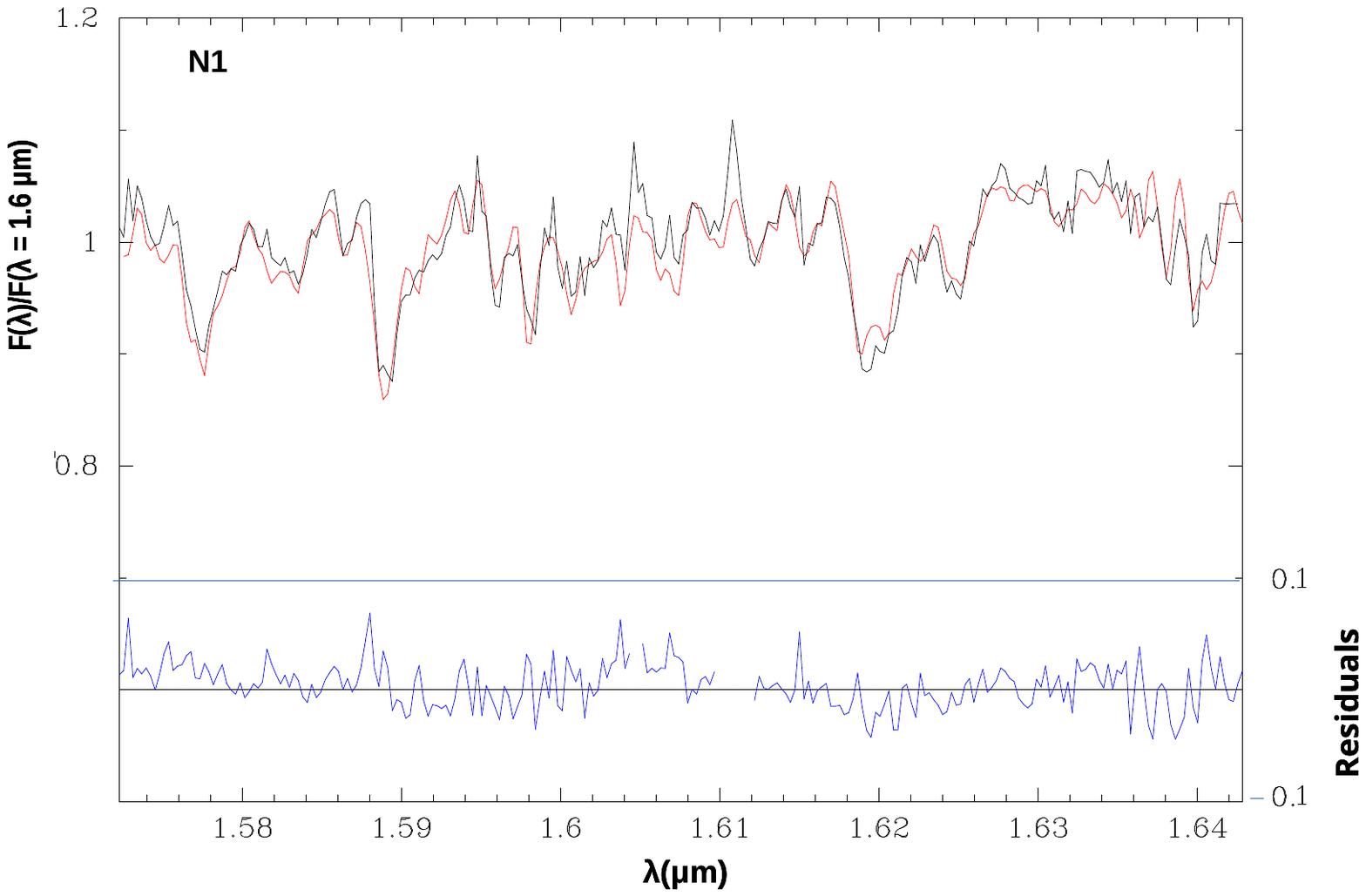}
\includegraphics[width=8.4cm]{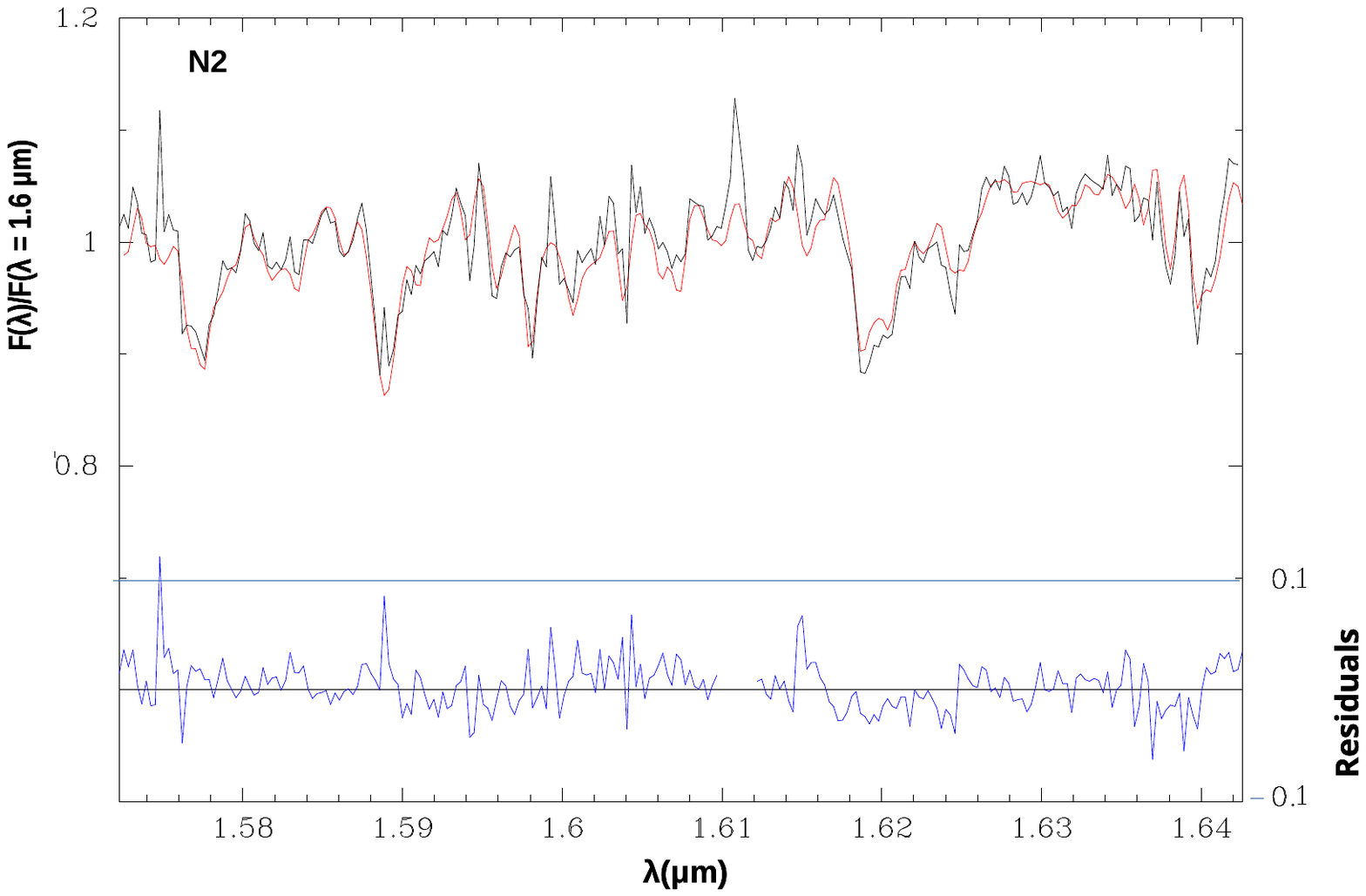}
\caption[]{Stellar population fits to the three regions of
  NGC~6221. From top to bottom: nucleus, N1 and N2 (see
  Fig.~\ref{fig:imN6221}).}
\label{fig:fitN6221}
\end{figure}

NGC~6221, is a barred Sbc galaxy located within a small galaxy group, with a bar clearly visible in the optical and IR. This X-ray bright Seyfert galaxy is known to exhibit clear signs of intense star formation together with a weak Seyfert nucleus, exhibiting both signatures of a Seyfert~1 in the X-rays and a typical reddened (Av=3) starburst-like spectrum in the optical \citep[e.g. ][]{Levenson+01, LaFranca+16}. 

The sharp-divided image of NGC~6221 is shown in Fig.~\ref{fig:imN6221}
with the studied regions superimposed. We study the nuclear spectrum within 80 pc, 
and  two regions towards the north up to 200 pc, excluding from our study the south conspicuous HII region spectrum \citep{Boisson+02}. The corresponding stellar populations are given in
Table~\ref{tab:n6221res}, with spectra and fits
shown in Fig.~\ref{fig:fitN6221}. We note that the Brackett (4-13) emission line 
at 1.61~$\mu$m visible on all spectra has been excluded from the fit.

Although the stellar  spectrum is strongly diluted (58\% power-law contribution) in the very nucleus, one can see in  Fig.~\ref{fig:fitN6221} conspicuous stellar features due 
to the large contribution of M supergiants which have strong narrow features. We find 60\% and 40\% contributions of old stars and very young stars respectively. This young component confirms the presence of a very strong starburst 
on top of the AGN.

The stellar population at $\sim 100$~pc (N1) is old with hints of past 
 star formation in a dusty environment suggested by the small intermediate age component. Further away, at $\sim 250$~pc  from the nucleus (N2), in an even more dusty environment, it appears that a succession of bursts have 
occurred. Both regions exhibit a mixture of stars that are rich and poor in metals.  We thus see a clear stellar population gradient together with a patchy extinction.
All this is in coherence with the composite classification of NGC~6221.

\subsection{NGC~7582}

\begin{table}
  \caption{Parameters of the spectral synthesis for NGC~7582. }
\begin{tabular}{rrrrr}
\hline
\hline
         &    S1 & S2             & N1        & N2\\
         & 40-120pc& 120-215pc  & 40-120pc &120-215pc\\
\hline
$\sigma$  km/s&  70  &  70      &  70      &  70  \\
\% stell &  100$\pm$0 &  100$\pm$0          & 100$\pm$0      & 100$\pm$0  \\
\% PL    &  0   &  0            &  0       & 0     \\
\%  BB      &  -   &  2$\pm$0          &  -       & -   \\
T (BB)   &  -   &  200$\pm$20     &  -       & -   \\
E(B-V)   &  -   &  -  &  0.52$\pm$0.17        & 0.78$\pm$0.02  \\
\hline
A0V      &         &        &  2$\pm$1            &    \\
F2V      & 58$\pm$1      & 44$\pm$2     & 25$\pm$1          & 35$\pm$1  \\
F8III    &         &        &  20$\pm$1          &   11$\pm$1  \\
K5-M0III &   2$\pm$2     & 24$\pm$2     &   18$\pm$1         &  41$\pm$1 \\
M4III    & 10$\pm$1      & 4$\pm$1      & 9$\pm$1            &     \\ 
rF8Ib    &         & 1$\pm$1      &              &     \\
rK4I      &         & 2$\pm$1      &              &    \\
M2I      & 30$\pm$1      & 25$\pm$1     & 26$\pm$1           &  13$\pm$1 \\
\hline
\end{tabular}
\\ 
\label{tab:n7582res}
\end{table}

\begin{figure}
\centering
\includegraphics[angle=0,width=7cm]{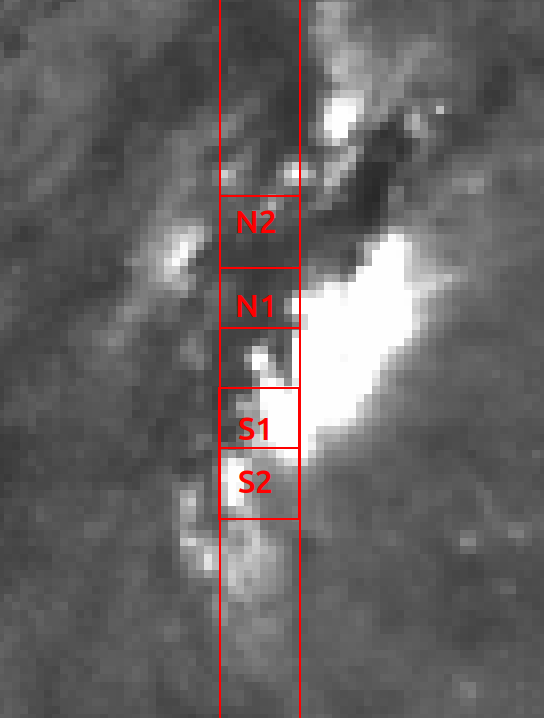}
\caption[]{Sharp-divided IR HST image in the F606WN filter of
  NGC~7582 with the ISAAC slit superimposed. The north (N1, N2) and south (S1, S2) regions are indicated.}
\label{fig:imN7582}
\end{figure}

\begin{figure}
\centering
\includegraphics[width=8.4cm]{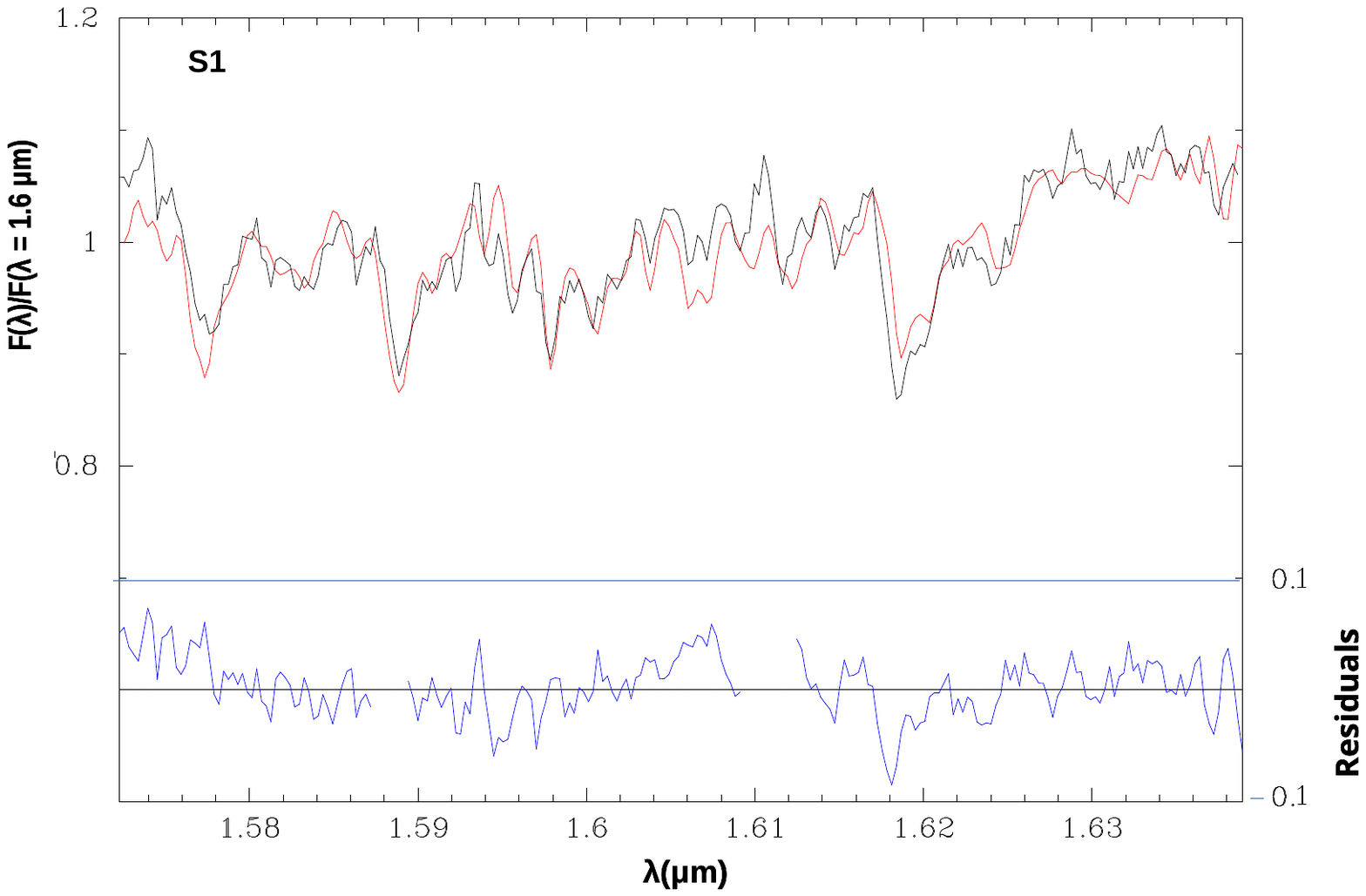}
\includegraphics[width=8.4cm]{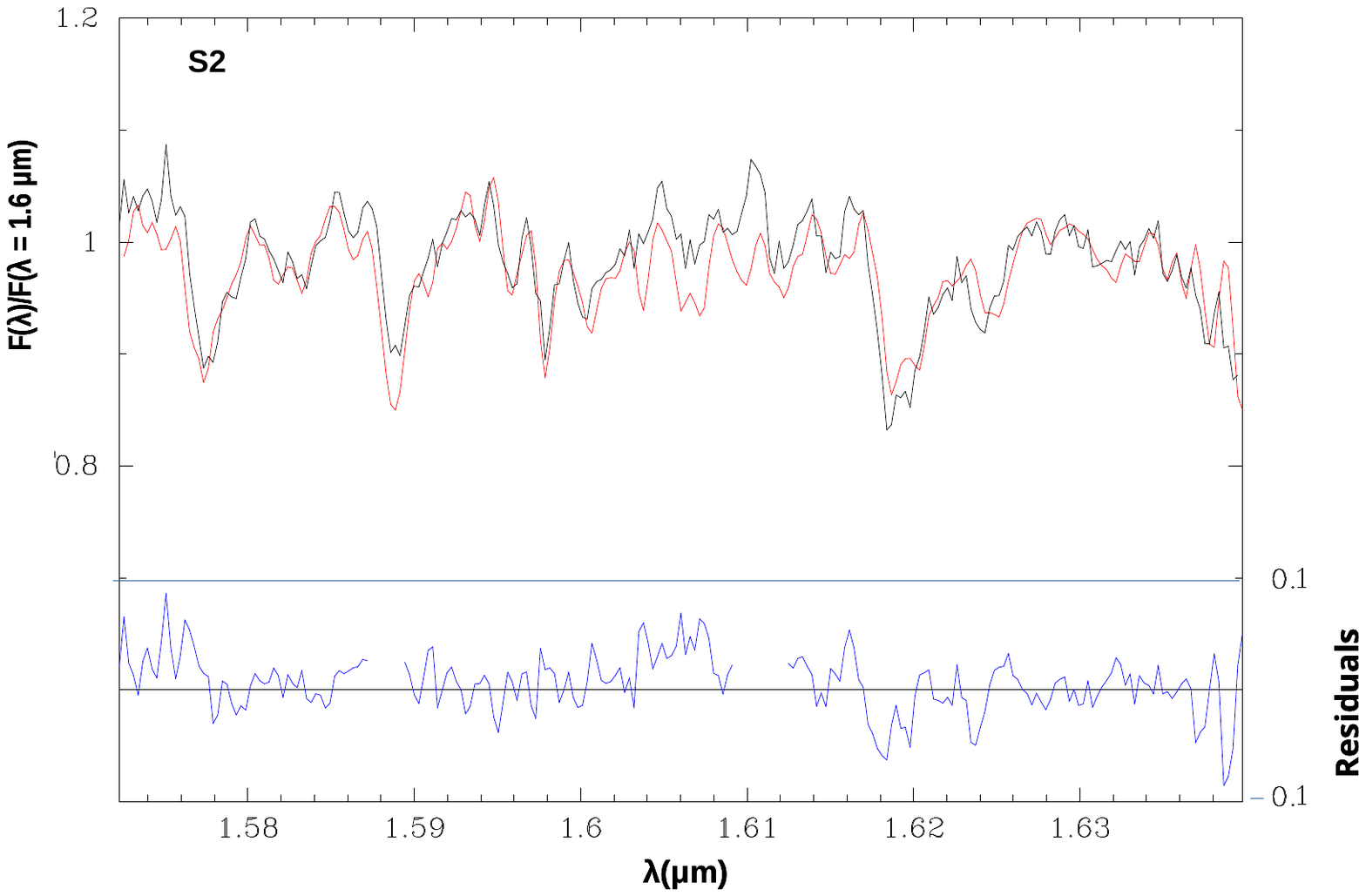}
\includegraphics[width=8.4cm]{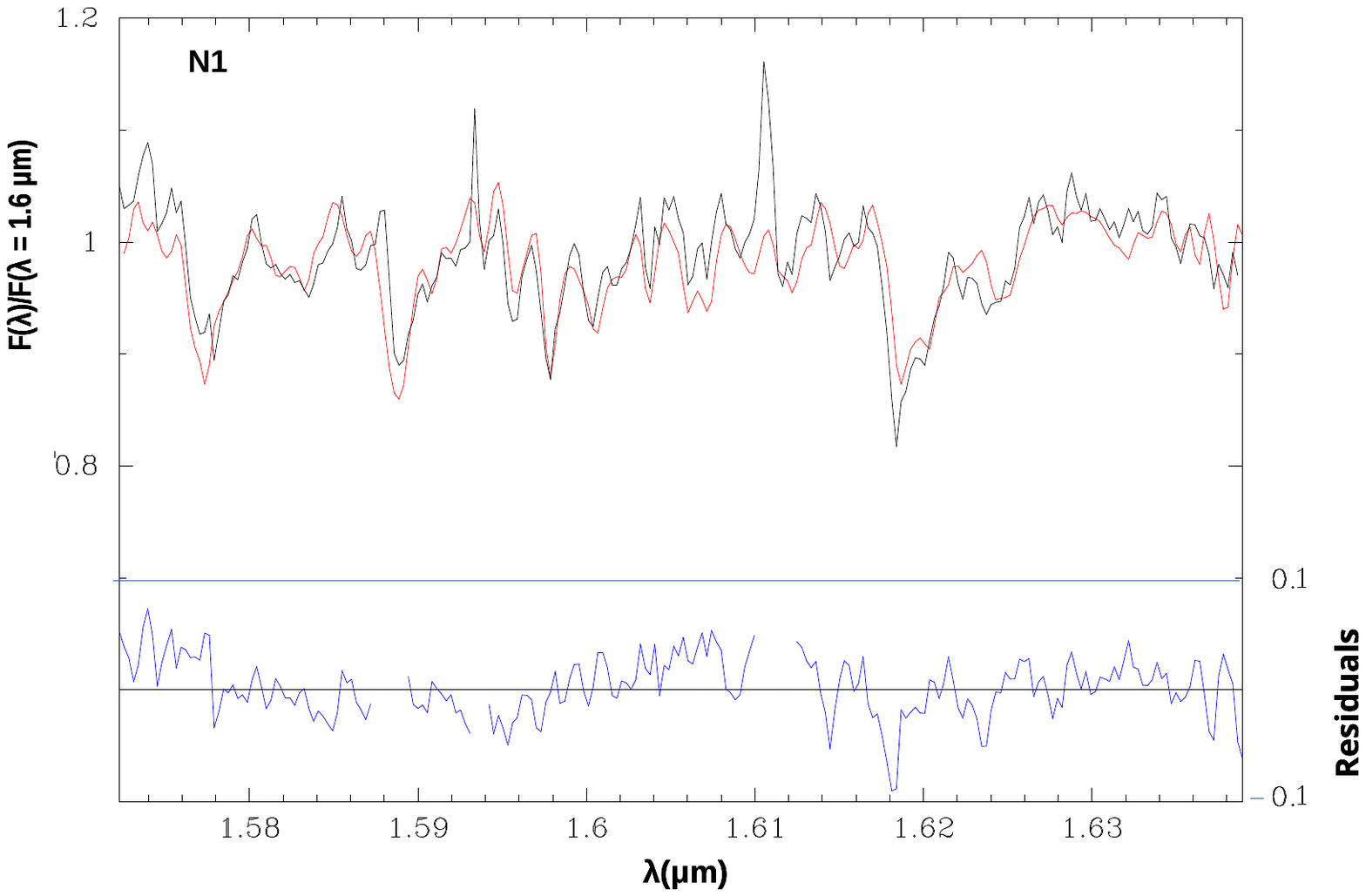}
\includegraphics[width=8.4cm]{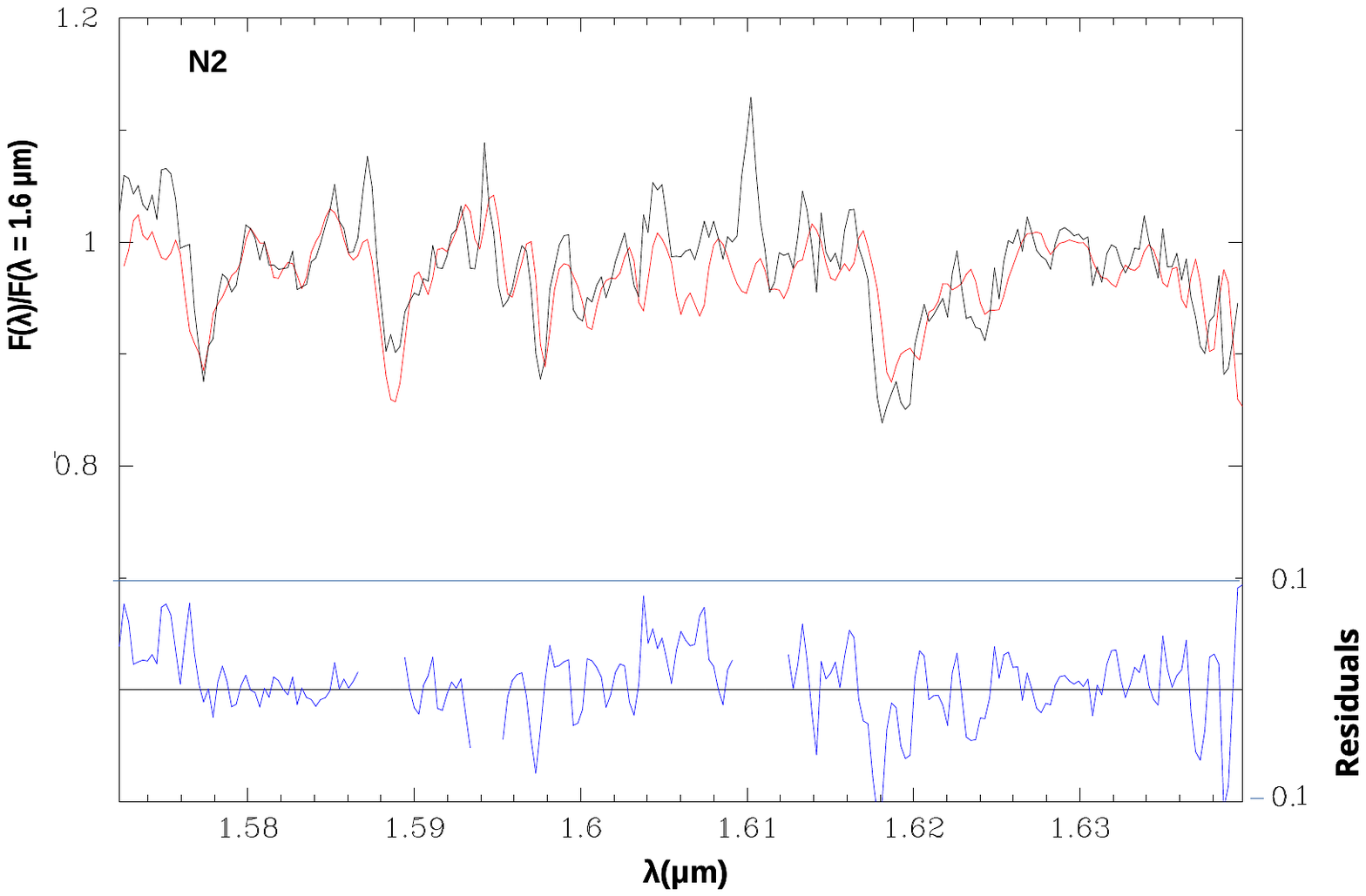}
\caption[]{Stellar population fits to the four regions of
  NGC~7582. From top to bottom: N2, N1, S1, and S2 (see Fig.~\ref{fig:imN7582}).}
\label{fig:fitN7582}
\end{figure}

NGC 7582 is a nearby barred galaxy in the Grus quartet that harbours a nuclear 
starburst surrounding the active nucleus \citep{Wold06,Bianchi+07}. The Galactic column density towards NGC~7582 is very low. This X-ray bright Seyfert 2 galaxy \citep{Bianchi+09} is a typical  ``changing look AGN'', a broad emission line component being reported by \cite{Aretxaga+99}. This component was initially proposed as caused by a supernova explosion, but was later confirmed to be from an AGN \citep{Ricci+18}. 

The sharp-divided image of NGC~7582 (Fig.~\ref{fig:imN7582}) shows a disturbed morphology with an obscured nucleus, and a dust lane crossing over the nuclear region. 
 Due to strong dilution by the AGN non-thermal continuum, 
the nuclear spectrum (80 arcsec diameter) is useless for stellar population 
studies. However, two external regions on each
 side of the nucleus were extracted, whose stellar populations are given in
Table~\ref{tab:n7582res}, with the corresponding spectra and fits
shown in Fig.~\ref{fig:fitN7582}. The northern regions are shining through  dust patches. This translates into higher E(B-V) in those regions than on the other side of the nucleus. We note that the Brackett (4-13) emission line 
at 1.61~$\mu$m visible on all spectra has been excluded from the fit.

We find an important population of evolved giant stars ($\sim (4-6) \times 10^9$ yrs), intermediate ones ($\sim 4 \times 10^8$ yrs) as well as very young (10$^{7}$ yrs) stars in the four selected regions. This is in line with what is found in the optical range  by \cite{Burtscher+21}, and in the near IR by  \cite{Riffel+24} within the 200 central parsecs.

 In contrast with the traditional view that Seyfert 2s are essentially composed of an old stellar population, we do not find any cold dwarf (i.e. stars older than 10$^{10}$ yrs)  whatever the region sampled. Our results (neither dwarf nor metal-rich stars)  points to a strong ongoing burst of star formation in the inner 200 pc. NGC~7582 may be at a star forming stage.

\section{Summary and conclusions}
\label{sec:discussion}

Taking advantage of the rich near-IR H-band, we investigated the stellar populations of the central few 100$~$pc of the host galaxy of seven active galactic nuclei, with the help of a stellar spectral database that allowed us to detect the young star contribution down to  $\leq 10^7$ yrs.

Our results can be summarised as follows:

First, in the inner 100~pc, we detect some metal-rich population in all AGNs but NGC~7582.  The only Seyfert~1 for which we could synthesise in detail the nucleus, MCG~06-30-15, shows a strong nuclear starburst. The two composite Seyfert~2s, NGC~6221 and NGC~7582, are also dominated by a starburst component, while no young stars are detected in the genuine Seyfert 2 galaxies, NGC~2110, and NGC~3185, nor in NGC~2992. 

Moreover, with our good spatial sampling, we could show how the star formation history of the inner nucleus of the galaxies is highly heterogeneous, as young bursts of star formation, intermediate (poststarburst) and old stellar populations all appear in widely varying proportions at all places. For example, one can note that in the four regions within 100~pc of MCG 06-30-15 (for which we have the best spatial and spectral resolutions), we find a large variety of populations. We could highlight such a diversity thanks to the IR domain, rich in stellar features, and where the star contribution is higher. Thus, there may not be only one process or relation between fuelling by star formation and power of the AGN.

Second, outside the nucleus ($>$100~pc) in both Seyfert 1 and Seyfert 2, we have resolved an intermediate or even young stellar population.

These results indicate that an active nucleus always seems to be linked to some past or even ongoing stellar formation.
This was also noted by, for example, \cite{Davies+07} or \cite{Riffel+22} who, based on high spatial resolution integral ﬁeld observations, found that the near-IR light in the inner few 100~pc of active galaxies is primarily composed on average of moderately young (a few $10^7$ to $10^8$ yrs) to intermediate-age (a few $10^8$ yrs) stellar populations, and that the age of the starburst seems connected to the AGN’s luminosity. However, those studies concern only the post-starburst population, with ages $\sim 10^8$ – $10^9$ yr.

Based on results of the stellar population study, in the optical, of a sample of 12 galaxies covering a range of morphology types and nuclear activities, \cite{Boisson+00} suggested the existence of a relationship between stellar population and activity type. The picture emerging from this study was that of an evolutionary sequence rather than a strict unified scheme for AGNs. Indeed, differences in the stellar populations within the nuclear regions were found to follow the degree of nuclear activity. Seyfert 2 galaxies appear to exhibit fossil star formation activity, that is, a stellar population younger than found in normal galaxies, but somewhat older and more metal-rich than found in starburst galaxies. In an optical study of the population synthesis of the inner 300~pc of star forming and active galaxies, including the three Seyfert 2 galaxies of our sample, \cite{Burtscher+21}  found a dominance of young stellar populations too, while this was not the case for their control sample of inactive galaxies. They also showed that such young populations are not necessarily indicative of ongoing star formation. This is along the lines of the evolutionary sequence of \cite{Boisson+00,Boisson+04}. Based on the similar nuclear star formation histories of the Seyfert 2s and the star forming control sample galaxies, \cite{Burtscher+21} further speculate that the starbursts probably turned into Seyferts for some fraction of their time.

In the present paper, in the IR, we find that indeed star-forming galaxies may turn into Seyfert 1s, while star formation in Seyfert 2s, if any, appears outside the nuclear region. We note that NGC 2992, for example, is known to change from type 1 to type~2.

In conclusion, while there is little doubt that the above observations indicate that AGN activity and star formation do often coincide within the very centre of a given galaxy, we find a great diversity of stellar populations, and no clear correlation with the level of activity of the central engine or with the galaxy morphological type, most probably linked to the fact that several of these objects vary from Seyfert 1 to Seyfert 2 and even LINER.

\begin{acknowledgements}

We thank the referee for comments and suggestions that helped us to improve the paper.
I.M. acknowledges financial support
from the State Agency for Research of the Spanish MCIU, through the “Center of Excellence Severo Ochoa” award to the Instituto de Astrofísica de Andalucía
(CEX2021-001131-S), and through PID2022-140871NB-C21.
F.D. acknowledges long-term support from CNES. 
This research has made use of the NASA/IPAC Infrared Science Archive, which is funded by1998 the National Aeronautics and Space 
Administration and operated by the California Institute of Technology, the NASA’s Astrophysics Data System bibliographic services, and of the Centre de Donn\'ees de Strasbourg. We gratefully thank Sandra Raimundo who provided us with the VLT-SINFONI spectra of MCG-6-30-15.

\end{acknowledgements}

\bibliographystyle{aa}
\bibliography{sample.bib}

\begin{thebibliography}{94}
\expandafter\ifx\csname natexlab\endcsname\relax\def\natexlab#1{#1}\fi

\bibitem[{{Adibekyan} {et~al.}(2012){Adibekyan}, {Sousa}, {Santos}, {Delgado
  Mena}, {Gonz{\'a}lez Hern{\'a}ndez}, {Israelian}, {Mayor}, \&
  {Khachatryan}}]{Adibekyan+12}
{Adibekyan}, V.~Z., {Sousa}, S.~G., {Santos}, N.~C., {et~al.} 2012, \aap, 545,
  A32

\bibitem[{{Af{\c{s}}ar} {et~al.}(2012){Af{\c{s}}ar}, {Sneden}, \&
  {For}}]{Afsar+12}
{Af{\c{s}}ar}, M., {Sneden}, C., \& {For}, B.~Q. 2012, \aj, 144, 20

\bibitem[{{Andrievsky} {et~al.}(2013){Andrievsky}, {L{\'e}pine}, {Korotin},
  {Luck}, {Kovtyukh}, \& {Maciel}}]{Andrievsky+13}
{Andrievsky}, S.~M., {L{\'e}pine}, J.~R.~D., {Korotin}, S.~A., {et~al.} 2013,
  \mnras, 428, 3252

\bibitem[{{Aretxaga} {et~al.}(1999){Aretxaga}, {Joguet}, {Kunth}, {Melnick}, \&
  {Terlevich}}]{Aretxaga+99}
{Aretxaga}, I., {Joguet}, B., {Kunth}, D., {Melnick}, J., \& {Terlevich}, R.~J.
  1999, \apjl, 519, L123

\bibitem[{{Beckmann} \& {Do Cao}(2010)}]{Beckmann2010}
{Beckmann}, V. \& {Do Cao}, O. 2010, in Eighth Integral Workshop. The Restless
  Gamma-ray Universe (INTEGRAL 2010), 81

\bibitem[{{Bentz} {et~al.}(2009){Bentz}, {Peterson}, {Netzer}, {Pogge}, \&
  {Vestergaard}}]{Bentz+09}
{Bentz}, M.~C., {Peterson}, B.~M., {Netzer}, H., {Pogge}, R.~W., \&
  {Vestergaard}, M. 2009, \apj, 697, 160

\bibitem[{{Bianchi} {et~al.}(2007){Bianchi}, {Chiaberge}, {Piconcelli}, \&
  {Guainazzi}}]{Bianchi+07}
{Bianchi}, S., {Chiaberge}, M., {Piconcelli}, E., \& {Guainazzi}, M. 2007,
  \mnras, 374, 697

\bibitem[{{Bianchi} {et~al.}(2009){Bianchi}, {Piconcelli}, {Chiaberge},
  {Bail{\'o}n}, {Matt}, \& {Fiore}}]{Bianchi+09}
{Bianchi}, S., {Piconcelli}, E., {Chiaberge}, M., {et~al.} 2009, \apj, 695, 781

\bibitem[{{Bica}(1988)}]{Bica1988}
{Bica}, E. 1988, \aap, 195, 76

\bibitem[{{Boesgaard} \& {Tripicco}(1987)}]{Boesgaard+87}
{Boesgaard}, A.~M. \& {Tripicco}, M.~J. 1987, \apj, 313, 389

\bibitem[{{Boisson} {et~al.}(2002){Boisson}, {Coup{\'e}}, {Cuby}, {Joly}, \&
  {Ward}}]{Boisson+02}
{Boisson}, C., {Coup{\'e}}, S., {Cuby}, J.~G., {Joly}, M., \& {Ward}, M.~J.
  2002, \aap, 396, 489

\bibitem[{{Boisson} {et~al.}(2000){Boisson}, {Joly}, {Moultaka}, {Pelat}, \&
  {Serote Roos}}]{Boisson+00}
{Boisson}, C., {Joly}, M., {Moultaka}, J., {Pelat}, D., \& {Serote Roos}, M.
  2000, \aap, 357, 850

\bibitem[{{Boisson} {et~al.}(2004){Boisson}, {Joly}, {Pelat}, \&
  {Ward}}]{Boisson+04}
{Boisson}, C., {Joly}, M., {Pelat}, D., \& {Ward}, M.~J. 2004, \aap, 428, 373

\bibitem[{{Bruzual} \& {Charlot}(2003)}]{Bruzual03}
{Bruzual}, G. \& {Charlot}, S. 2003, \mnras, 344, 1000

\bibitem[{{Burtscher} {et~al.}(2021){Burtscher}, {Davies}, {Shimizu}, {Riffel},
  {Rosario}, {Hicks}, {Lin}, {Riffel}, {Schartmann}, {Schnorr-M{\"u}ller},
  {Storchi-Bergmann}, {Orban de Xivry}, \& {Veilleux}}]{Burtscher+21}
{Burtscher}, L., {Davies}, R.~I., {Shimizu}, T.~T., {et~al.} 2021, \aap, 654,
  A132

\bibitem[{{Carrillo} {et~al.}(1999){Carrillo}, {Masegosa}, {Dultzin-Hacyan}, \&
  {Ordo{\~n}ez}}]{Carrillo+99}
{Carrillo}, R., {Masegosa}, J., {Dultzin-Hacyan}, D., \& {Ordo{\~n}ez}, R.
  1999, \rmxaa, 35, 187

\bibitem[{{Cesetti} {et~al.}(2009){Cesetti}, {Ivanov}, {Morelli}, {Pizzella},
  {Buson}, {Corsini}, {Dalla Bont{\`a}}, {Stiavelli}, \&
  {Bertola}}]{Cesetti2009}
{Cesetti}, M., {Ivanov}, V.~D., {Morelli}, L., {et~al.} 2009, \aap, 497, 41

\bibitem[{{Chapman} {et~al.}(2000){Chapman}, {Morris}, {Alonso-Herrero}, \&
  {Falcke}}]{Chapman2000}
{Chapman}, S.~C., {Morris}, S.~L., {Alonso-Herrero}, A., \& {Falcke}, H. 2000,
  \mnras, 314, 263

\bibitem[{{Cid Fernandes}(2018)}]{Cid2018}
{Cid Fernandes}, R. 2018, \mnras, 480, 4480

\bibitem[{{Cid Fernandes} {et~al.}(2005){Cid Fernandes}, {Mateus}, {Sodr{\'e}},
  {Stasi{\'n}ska}, \& {Gomes}}]{Cid+05}
{Cid Fernandes}, R., {Mateus}, A., {Sodr{\'e}}, L., {Stasi{\'n}ska}, G., \&
  {Gomes}, J.~M. 2005, \mnras, 358, 363

\bibitem[{{Dahmer-Hahn} {et~al.}(2022){Dahmer-Hahn}, {Riffel},
  {Rodr{\'\i}guez-Ardila}, {Riffel}, {Storchi-Bergmann}, {Marinello}, {Davies},
  {Burtscher}, {Ruschel-Dutra}, \& {Rosario}}]{Dahmer+22}
{Dahmer-Hahn}, L.~G., {Riffel}, R., {Rodr{\'\i}guez-Ardila}, A., {et~al.} 2022,
  \mnras, 509, 4653

\bibitem[{{Dallier} {et~al.}(1996){Dallier}, {Boisson}, \&
  {Joly}}]{Dallier1996}
{Dallier}, R., {Boisson}, C., \& {Joly}, M. 1996, \aaps, 116, 239

\bibitem[{{Dametto} {et~al.}(2019){Dametto}, {Riffel}, {Colina}, {Riffel},
  {Piqueras L{\'o}pez}, {Davies}, {Burtscher}, {Menezes}, {Arribas},
  {Pastoriza}, {Labiano}, {Storchi-Bergmann}, {Dahmer-Hahn}, \&
  {Sales}}]{Dametto+19}
{Dametto}, N.~Z., {Riffel}, R., {Colina}, L., {et~al.} 2019, \mnras, 482, 4437

\bibitem[{{Davies} {et~al.}(2007){Davies}, {M{\"u}ller S{\'a}nchez}, {Genzel},
  {Tacconi}, {Hicks}, {Friedrich}, \& {Sternberg}}]{Davies+07}
{Davies}, R.~I., {M{\"u}ller S{\'a}nchez}, F., {Genzel}, R., {et~al.} 2007,
  \apj, 671, 1388

\bibitem[{{Diniz} {et~al.}(2017){Diniz}, {Riffel}, {Riffel}, {Crenshaw},
  {Storchi-Bergmann}, {Fischer}, {Schmitt}, \& {Kraemer}}]{Diniz+17}
{Diniz}, M.~R., {Riffel}, R.~A., {Riffel}, R., {et~al.} 2017, \mnras, 469, 3286

\bibitem[{{Diniz} {et~al.}(2019){Diniz}, {Riffel}, {Storchi-Bergmann}, \&
  {Riffel}}]{Diniz+19}
{Diniz}, M.~R., {Riffel}, R.~A., {Storchi-Bergmann}, T., \& {Riffel}, R. 2019,
  \mnras, 487, 3958

\bibitem[{{Edvardsson} {et~al.}(1993){Edvardsson}, {Andersen}, {Gustafsson},
  {Lambert}, {Nissen}, \& {Tomkin}}]{Edvardsson+93}
{Edvardsson}, B., {Andersen}, J., {Gustafsson}, B., {et~al.} 1993, \aap, 275,
  101

\bibitem[{{Esquej} {et~al.}(2014){Esquej}, {Alonso-Herrero},
  {Gonz{\'a}lez-Mart{\'\i}n}, {H{\"o}nig}, {Hern{\'a}n-Caballero}, {Roche},
  {Ramos Almeida}, {Mason}, {D{\'\i}az-Santos}, {Levenson}, {Aretxaga},
  {Rodr{\'\i}guez Espinosa}, \& {Packham}}]{Esquej+14}
{Esquej}, P., {Alonso-Herrero}, A., {Gonz{\'a}lez-Mart{\'\i}n}, O., {et~al.}
  2014, \apj, 780, 86

\bibitem[{{Evans} {et~al.}(2006){Evans}, {Lee}, {Kamenetska}, {Gallagher},
  {Kraft}, {Hardcastle}, \& {Weaver}}]{Evans+06}
{Evans}, D.~A., {Lee}, J.~C., {Kamenetska}, M., {et~al.} 2006, \apj, 653, 1121

\bibitem[{{Faber}(1972)}]{Faber_1972}
{Faber}, S.~M. 1972, \aap, 20, 361

\bibitem[{{Ferruit} {et~al.}(2000){Ferruit}, {Wilson}, \&
  {Mulchaey}}]{Ferruit00}
{Ferruit}, P., {Wilson}, A.~S., \& {Mulchaey}, J. 2000, \apjs, 128, 139

\bibitem[{{Filippenko} \& {Sargent}(1985)}]{Filippenko85}
{Filippenko}, A.~V. \& {Sargent}, W.~L.~W. 1985, \apjs, 57, 503

\bibitem[{{Fr{\'e}maux} {et~al.}(2006){Fr{\'e}maux}, {Kupka}, {Boisson},
  {Joly}, \& {Tsymbal}}]{Fremaux+06}
{Fr{\'e}maux}, J., {Kupka}, F., {Boisson}, C., {Joly}, M., \& {Tsymbal}, V.
  2006, \aap, 449, 109

\bibitem[{{Fr{\'e}maux} {et~al.}(2007){Fr{\'e}maux}, {Pelat}, {Boisson}, \&
  {Joly}}]{Fremaux+07}
{Fr{\'e}maux}, J., {Pelat}, D., {Boisson}, C., \& {Joly}, M. 2007, \aap, 473,
  377

\bibitem[{{Friedrich} {et~al.}(2010){Friedrich}, {Davies}, {Hicks}, {Engel},
  {M{\"u}ller-S{\'a}nchez}, {Genzel}, \& {Tacconi}}]{Friedrich+10}
{Friedrich}, S., {Davies}, R.~I., {Hicks}, E.~K.~S., {et~al.} 2010, \aap, 519,
  A79

\bibitem[{{Garc{\'\i}a-Bernete} {et~al.}(2015){Garc{\'\i}a-Bernete}, {Ramos
  Almeida}, {Acosta-Pulido}, {Alonso-Herrero}, {S{\'a}nchez-Portal},
  {Castillo}, {Pereira-Santaella}, {Esquej}, {Gonz{\'a}lez-Mart{\'\i}n},
  {D{\'\i}az-Santos}, {Roche}, {Fisher}, {Povi{\'c}}, {P{\'e}rez Garc{\'\i}a},
  {Valtchanov}, {Packham}, \& {Levenson}}]{GarciaBernete+15}
{Garc{\'\i}a-Bernete}, I., {Ramos Almeida}, C., {Acosta-Pulido}, J.~A.,
  {et~al.} 2015, \mnras, 449, 1309

\bibitem[{{Gebhardt} {et~al.}(2000){Gebhardt}, {Bender}, {Bower}, {Dressler},
  {Faber}, {Filippenko}, {Green}, {Grillmair}, {Ho}, {Kormendy}, {Lauer},
  {Magorrian}, {Pinkney}, {Richstone}, \& {Tremaine}}]{Gebhardt2000}
{Gebhardt}, K., {Bender}, R., {Bower}, G., {et~al.} 2000, \apjl, 539, L13

\bibitem[{{Gomes} \& {Papaderos}(2017)}]{Gomes2017}
{Gomes}, J.~M. \& {Papaderos}, P. 2017, \aap, 603, A63

\bibitem[{{Gon{\c{c}}alves} {et~al.}(1999){Gon{\c{c}}alves}, {V{\'e}ron-Cetty},
  \& {V{\'e}ron}}]{Goncalves_1999}
{Gon{\c{c}}alves}, A.~C., {V{\'e}ron-Cetty}, M.~P., \& {V{\'e}ron}, P. 1999,
  \aaps, 135, 437

\bibitem[{{Guolo-Pereira} {et~al.}(2021){Guolo-Pereira}, {Ruschel-Dutra},
  {Storchi-Bergmann}, {Schnorr-M{\"u}ller}, {Cid Fernandes}, {Couto},
  {Dametto}, \& {Hernandez-Jimenez}}]{Guolo-Pereira2021+}
{Guolo-Pereira}, M., {Ruschel-Dutra}, D., {Storchi-Bergmann}, T., {et~al.}
  2021, \mnras, 502, 3618

\bibitem[{{Hauck} {et~al.}(1985){Hauck}, {Foy}, \& {Proust}}]{Hauck+85}
{Hauck}, B., {Foy}, R., \& {Proust}, D. 1985, \aap, 149, 167

\bibitem[{{Ivanov} {et~al.}(2004){Ivanov}, {Rieke}, {Engelbracht},
  {Alonso-Herrero}, {Rieke}, \& {Luhman}}]{Ivanov2004}
{Ivanov}, V.~D., {Rieke}, M.~J., {Engelbracht}, C.~W., {et~al.} 2004, \apjs,
  151, 387

\bibitem[{{James} \& {Percival}(2016)}]{James+16}
{James}, P.~A. \& {Percival}, S.~M. 2016, \mnras, 457, 917

\bibitem[{{Koleva} \& {Vazdekis}(2012)}]{Koleva+12}
{Koleva}, M. \& {Vazdekis}, A. 2012, \aap, 538, A143

\bibitem[{{Kormendy} \& {Ho}(2013)}]{Kormendy2013}
{Kormendy}, J. \& {Ho}, L.~C. 2013, \araa, 51, 511

\bibitem[{{La Franca} {et~al.}(2016){La Franca}, {Onori}, {Ricci}, {Bianchi},
  {Marconi}, {Sani}, \& {Vignali}}]{LaFranca+16}
{La Franca}, F., {Onori}, F., {Ricci}, F., {et~al.} 2016, Frontiers in
  Astronomy and Space Sciences, 3, 12

\bibitem[{{Lanz} {et~al.}(2013){Lanz}, {Zezas}, {Brassington}, {Smith},
  {Ashby}, {da Cunha}, {Fazio}, {Hayward}, {Hernquist}, \& {Jonsson}}]{Lanz+13}
{Lanz}, L., {Zezas}, A., {Brassington}, N., {et~al.} 2013, \apj, 768, 90

\bibitem[{{Le Borgne} {et~al.}(2004){Le Borgne}, {Rocca-Volmerange},
  {Prugniel}, {Lan{\c{c}}on}, {Fioc}, \& {Soubiran}}]{leborgne04}
{Le Borgne}, D., {Rocca-Volmerange}, B., {Prugniel}, P., {et~al.} 2004, \aap,
  425, 881

\bibitem[{{Lee} {et~al.}(2011){Lee}, {Beers}, {Allende Prieto}, {Lai},
  {Rockosi}, {Morrison}, {Johnson}, {An}, {Sivarani}, \& {Yanny}}]{Lee+11}
{Lee}, Y.~S., {Beers}, T.~C., {Allende Prieto}, C., {et~al.} 2011, \aj, 141, 90

\bibitem[{{Levenson} {et~al.}(2001){Levenson}, {Cid Fernandes}, {Weaver},
  {Heckman}, \& {Storchi-Bergmann}}]{Levenson+01}
{Levenson}, N.~A., {Cid Fernandes}, R., J., {Weaver}, K.~A., {Heckman}, T.~M.,
  \& {Storchi-Bergmann}, T. 2001, \apj, 557, 54

\bibitem[{{Luck}(2014)}]{Luck14}
{Luck}, R.~E. 2014, \aj, 147, 137

\bibitem[{{Mallmann} {et~al.}(2018){Mallmann}, {Riffel}, {Storchi-Bergmann},
  {Rembold}, {Riffel}, {Schimoia}, {da Costa}, {{\'A}vila-Reese}, {Sanchez},
  {Machado}, {Cirolini}, {Ilha}, \& {Nascimento}}]{Mallmann+18}
{Mallmann}, N.~D., {Riffel}, R., {Storchi-Bergmann}, T., {et~al.} 2018, \mnras,
  478, 5491

\bibitem[{{Mann} {et~al.}(2015){Mann}, {Feiden}, {Gaidos}, {Boyajian}, \& {von
  Braun}}]{Mann+15}
{Mann}, A.~W., {Feiden}, G.~A., {Gaidos}, E., {Boyajian}, T., \& {von Braun},
  K. 2015, \apj, 804, 64

\bibitem[{{M{\'a}rquez} {et~al.}(1998){M{\'a}rquez}, {Boisson}, {Durret}, \&
  {Petitjean}}]{Marquez+98}
{M{\'a}rquez}, I., {Boisson}, C., {Durret}, F., \& {Petitjean}, P. 1998, \aap,
  333, 459

\bibitem[{{M{\'a}rquez} {et~al.}(2003){M{\'a}rquez}, {Masegosa}, {Durret},
  {Gonz{\'a}lez Delgado}, {Moles}, {Maza}, {P{\'e}rez}, \& {Roth}}]{Marquez03}
{M{\'a}rquez}, I., {Masegosa}, J., {Durret}, F., {et~al.} 2003, \aap, 409, 459

\bibitem[{{Massarotti} {et~al.}(2008){Massarotti}, {Latham}, {Stefanik}, \&
  {Fogel}}]{Massarotti+08}
{Massarotti}, A., {Latham}, D.~W., {Stefanik}, R.~P., \& {Fogel}, J. 2008, \aj,
  135, 209

\bibitem[{{Meyer} {et~al.}(1998){Meyer}, {Edwards}, {Hinkle}, \&
  {Strom}}]{Meyer+98}
{Meyer}, M.~R., {Edwards}, S., {Hinkle}, K.~H., \& {Strom}, S.~E. 1998, \apj,
  508, 397

\bibitem[{{Moultaka} {et~al.}(2004){Moultaka}, {Boisson}, {Joly}, \&
  {Pelat}}]{Moultaka2004}
{Moultaka}, J., {Boisson}, C., {Joly}, M., \& {Pelat}, D. 2004, \aap, 420, 459

\bibitem[{{Moultaka} \& {Pelat}(2000)}]{Moultaka_Pelat2000}
{Moultaka}, J. \& {Pelat}, D. 2000, \mnras, 314, 409

\bibitem[{{Oliva} {et~al.}(1999){Oliva}, {Origlia}, {Maiolino}, \&
  {Moorwood}}]{Oliva+99}
{Oliva}, E., {Origlia}, L., {Maiolino}, R., \& {Moorwood}, A.~F.~M. 1999, \aap,
  350, 9

\bibitem[{{Origlia} {et~al.}(1993){Origlia}, {Moorwood}, \&
  {Oliva}}]{Origlia1993}
{Origlia}, L., {Moorwood}, A.~F.~M., \& {Oliva}, E. 1993, \aap, 280, 536

\bibitem[{{Pelat}(1997)}]{Pelat97}
{Pelat}, D. 1997, \mnras, 284, 365

\bibitem[{{Pelat}(1998)}]{Pelat98}
{Pelat}, D. 1998, \mnras, 299, 877

\bibitem[{{Pickles}(1985)}]{Pickles_1985}
{Pickles}, A.~J. 1985, \apj, 296, 340

\bibitem[{{Prugniel} {et~al.}(2011){Prugniel}, {Vauglin}, \&
  {Koleva}}]{Prugniel+11}
{Prugniel}, P., {Vauglin}, I., \& {Koleva}, M. 2011, \aap, 531, A165

\bibitem[{{Raimundo} {et~al.}(2017){Raimundo}, {Davies}, {Canning}, {Celotti},
  {Fabian}, \& {Gandhi}}]{Raimundo+17}
{Raimundo}, S.~I., {Davies}, R.~I., {Canning}, R.~E.~A., {et~al.} 2017, \mnras,
  464, 4227

\bibitem[{{Raimundo} {et~al.}(2013){Raimundo}, {Davies}, {Gandhi}, {Fabian},
  {Canning}, \& {Ivanov}}]{Raimundo+13}
{Raimundo}, S.~I., {Davies}, R.~I., {Gandhi}, P., {et~al.} 2013, \mnras, 431,
  2294

\bibitem[{{Ram{\'\i}rez} {et~al.}(2013){Ram{\'\i}rez}, {Allende Prieto}, \&
  {Lambert}}]{Ramirez+13}
{Ram{\'\i}rez}, I., {Allende Prieto}, C., \& {Lambert}, D.~L. 2013, \apj, 764,
  78

\bibitem[{{Rayner} {et~al.}(2009){Rayner}, {Cushing}, \& {Vacca}}]{Rayner_2009}
{Rayner}, J.~T., {Cushing}, M.~C., \& {Vacca}, W.~D. 2009, \apjs, 185, 289

\bibitem[{{Ricci} {et~al.}(2018){Ricci}, {Steiner}, {May}, {Garcia-Rissmann},
  \& {Menezes}}]{Ricci+18}
{Ricci}, T.~V., {Steiner}, J.~E., {May}, D., {Garcia-Rissmann}, A., \&
  {Menezes}, R.~B. 2018, \mnras, 473, 5334

\bibitem[{{Riffel} {et~al.}(2022){Riffel}, {Dahmer-Hahn}, {Riffel},
  {Storchi-Bergmann}, {Dametto}, {Davies}, {Burtscher}, {Bianchin},
  {Ruschel-Dutra}, {Ricci}, \& {Rosario}}]{Riffel+22}
{Riffel}, R., {Dahmer-Hahn}, L.~G., {Riffel}, R.~A., {et~al.} 2022, \mnras,
  512, 3906

\bibitem[{{Riffel} {et~al.}(2024){Riffel}, {Dahmer-Hahn}, {Vazdekis}, {Davies},
  {Rosario}, {Ramos Almeida}, {Audibert}, {Mart{\'\i}n-Navarro}, {Pires
  Martins}, {Rodr{\'\i}guez-Ardila}, {Riffel}, {Storchi-Bergmann},
  {Bertoldo-Coelho}, {Trevisan}, {Hicks}, {M{\"u}ller}, {Marinho}, \&
  {Veilleux}}]{Riffel+24}
{Riffel}, R., {Dahmer-Hahn}, L.~G., {Vazdekis}, A., {et~al.} 2024, \mnras, 531,
  554

\bibitem[{{Schlafly} \& {Finkbeiner}(2011)}]{SchlaflyFinkbeiner11}
{Schlafly}, E.~F. \& {Finkbeiner}, D.~P. 2011, \apj, 737, 103

\bibitem[{{Schnorr-M{\"u}ller} {et~al.}(2014){Schnorr-M{\"u}ller},
  {Storchi-Bergmann}, {Nagar}, {Robinson}, {Lena}, {Riffel}, \&
  {Couto}}]{SchnorrMuller+14}
{Schnorr-M{\"u}ller}, A., {Storchi-Bergmann}, T., {Nagar}, N.~M., {et~al.}
  2014, \mnras, 437, 1708

\bibitem[{{Scoville} {et~al.}(2023){Scoville}, {Faisst}, {Weaver}, {Toft},
  {McCracken}, {Ilbert}, {Diaz-Santos}, {Staguhn}, {Koda}, {Casey}, {Sanders},
  {Mobasher}, {Chartab}, {Sattari}, {Capak}, {Vanden Bout}, {Bongiorno},
  {Vlahakis}, {Sheth}, {Yun}, {Aussel}, {Laigle}, \& {Masters}}]{Scoville2023}
{Scoville}, N., {Faisst}, A., {Weaver}, J., {et~al.} 2023, \apj, 943, 82

\bibitem[{{Sheffield} {et~al.}(2012){Sheffield}, {Majewski}, {Johnston},
  {Cunha}, {Smith}, {Cheung}, {Hampton}, {David}, {Wagner-Kaiser}, {Johnson},
  {Kaplan}, {Miller}, \& {Patterson}}]{Sheffield+12}
{Sheffield}, A.~A., {Majewski}, S.~R., {Johnston}, K.~V., {et~al.} 2012, \apj,
  761, 161

\bibitem[{{Smith} \& {Lambert}(1986)}]{Smith+86}
{Smith}, V.~V. \& {Lambert}, D.~L. 1986, \apj, 311, 843

\bibitem[{{Stephens} \& {Frogel}(2004)}]{Stephens2004}
{Stephens}, A.~W. \& {Frogel}, J.~A. 2004, \aj, 127, 925

\bibitem[{{Stoklasov{\'a}} {et~al.}(2009){Stoklasov{\'a}}, {Ferruit},
  {Emsellem}, {Jungwiert}, {P{\'e}contal}, \& {S{\'a}nchez}}]{Stoklasova2009}
{Stoklasov{\'a}}, I., {Ferruit}, P., {Emsellem}, E., {et~al.} 2009, \aap, 500,
  1287

\bibitem[{{Storchi-Bergmann} {et~al.}(2001){Storchi-Bergmann}, {Gonz{\'a}lez
  Delgado}, {Schmitt}, {Cid Fernandes}, \& {Heckman}}]{StorchiB2001}
{Storchi-Bergmann}, T., {Gonz{\'a}lez Delgado}, R.~M., {Schmitt}, H.~R., {Cid
  Fernandes}, R., \& {Heckman}, T. 2001, \apj, 559, 147

\bibitem[{{Terlevich} {et~al.}(1990){Terlevich}, {Diaz}, \&
  {Terlevich}}]{Terlevich1990}
{Terlevich}, E., {Diaz}, A.~I., \& {Terlevich}, R. 1990, \mnras, 242, 271

\bibitem[{{Thomas} {et~al.}(2018){Thomas}, {Dopita}, {Kewley}, {Groves},
  {Sutherland}, {Hopkins}, \& {Blanc}}]{Thomas2018}
{Thomas}, A.~D., {Dopita}, M.~A., {Kewley}, L.~J., {et~al.} 2018, \apj, 856, 89

\bibitem[{{Tinsley}(1972)}]{Tinsley1972}
{Tinsley}, B.~M. 1972, \apj, 178, 319

\bibitem[{{Trippe} {et~al.}(2008){Trippe}, {Crenshaw}, {Deo}, \&
  {Dietrich}}]{Trippe+08}
{Trippe}, M.~L., {Crenshaw}, D.~M., {Deo}, R., \& {Dietrich}, M. 2008, \aj,
  135, 2048

\bibitem[{{Valenti} \& {Fischer}(2005)}]{Valenti+05}
{Valenti}, J.~A. \& {Fischer}, D.~A. 2005, \apjs, 159, 141

\bibitem[{{Vanden Berk} {et~al.}(2001){Vanden Berk}, {Richards}, {Bauer},
  {Strauss}, {Schneider}, {Heckman}, {York}, {Hall}, {Fan}, {Knapp},
  {Anderson}, {Annis}, {Bahcall}, {Bernardi}, {Briggs}, {Brinkmann}, {Brunner},
  {Burles}, {Carey}, {Castander}, {Connolly}, {Crocker}, {Csabai}, {Doi},
  {Finkbeiner}, {Friedman}, {Frieman}, {Fukugita}, {Gunn}, {Hennessy},
  {Ivezi{\'c}}, {Kent}, {Kunszt}, {Lamb}, {Leger}, {Long}, {Loveday}, {Lupton},
  {Meiksin}, {Merelli}, {Munn}, {Newberg}, {Newcomb}, {Nichol}, {Owen}, {Pier},
  {Pope}, {Rockosi}, {Schlegel}, {Siegmund}, {Smee}, {Snir}, {Stoughton},
  {Stubbs}, {SubbaRao}, {Szalay}, {Szokoly}, {Tremonti}, {Uomoto}, {Waddell},
  {Yanny}, \& {Zheng}}]{VandenBerk+01}
{Vanden Berk}, D.~E., {Richards}, G.~T., {Bauer}, A., {et~al.} 2001, \aj, 122,
  549

\bibitem[{{V{\'a}zquez} \& {Leitherer}(2005)}]{Vazquez05}
{V{\'a}zquez}, G.~A. \& {Leitherer}, C. 2005, \apj, 621, 695

\bibitem[{{Venn}(1995)}]{Venn95}
{Venn}, K.~A. 1995, \apjs, 99, 659

\bibitem[{{Vergely} {et~al.}(2002){Vergely}, {Lan{\c{c}}on}, \&
  {Mouhcine}}]{Vergely2002}
{Vergely}, J.~L., {Lan{\c{c}}on}, A., \& {Mouhcine}. 2002, \aap, 394, 807

\bibitem[{{Verro} {et~al.}(2022){Verro}, {Trager}, {Peletier}, {Lan{\c{c}}on},
  {Gonneau}, {Vazdekis}, {Prugniel}, {Chen}, {Coelho},
  {S{\'a}nchez-Bl{\'a}zquez}, {Martins}, {Arentsen}, {Lyubenova},
  {Falc{\'o}n-Barroso}, \& {Dries}}]{Verro_2022}
{Verro}, K., {Trager}, S.~C., {Peletier}, R.~F., {et~al.} 2022, \aap, 660, A34

\bibitem[{{Villaume} {et~al.}(2017){Villaume}, {Conroy}, {Johnson}, {Rayner},
  {Mann}, \& {van Dokkum}}]{Villaume_2017}
{Villaume}, A., {Conroy}, C., {Johnson}, B., {et~al.} 2017, \apjs, 230, 23

\bibitem[{{Ward} {et~al.}(1980){Ward}, {Penston}, {Blades}, \&
  {Turtle}}]{Ward+80}
{Ward}, M., {Penston}, M.~V., {Blades}, J.~C., \& {Turtle}, A.~J. 1980, \mnras,
  193, 563

\bibitem[{{Wilkinson} {et~al.}(2017){Wilkinson}, {Maraston}, {Goddard},
  {Thomas}, \& {Parikh}}]{Wilkinson+17}
{Wilkinson}, D.~M., {Maraston}, C., {Goddard}, D., {Thomas}, D., \& {Parikh},
  T. 2017, \mnras, 472, 4297

\bibitem[{{Wold} \& {Galliano}(2006)}]{Wold06}
{Wold}, M. \& {Galliano}, E. 2006, \mnras, 369, L47

\end{thebibliography}

\end{document}